\documentclass[aps,amsmath,amssymb,notitlepage,twocolumn,nofootinbib,floatfix]{revtex4-1}
\usepackage{amsthm}
\usepackage{amsfonts}
\usepackage{amsmath}
\usepackage{amssymb}
\usepackage{graphicx}
\usepackage{verbatim}
\usepackage{siunitx}
\usepackage{color}
\usepackage{textcomp}
\usepackage{multirow}
\usepackage{xcolor}
\usepackage{tikz}
\usepackage{pgfplots}
\usepackage{pdfpages}
\usepackage{hyperref}
\pgfplotsset{compat=newest}
\usetikzlibrary{external}
\newcommand{\fH}{\mathcal{H}}

\newlength\figureheight
\newlength\figurewidth
\renewcommand{\epsilon}{\varepsilon}

\begin{document}
\title{Performance of a quantum annealer on range-limited constraint satisfaction problems}
\author{A.~D.~King}\email[]{aking@dwavesys.com}
\author{T.~Lanting}
\author{R.~Harris}
\affiliation{D-Wave Systems Inc., 3033 Beta Avenue, Burnaby, British Columbia, Canada V5G 4M9}
\date{\today}

\begin{abstract}
The performance of a D-Wave Vesuvius quantum annealing processor was recently compared to a suite of classical algorithms on a class of constraint satisfaction instances based on frustrated loops.  However, the construction of these instances leads the maximum coupling strength to increase with problem size.  As a result, larger instances are subject to amplified analog control error, and are effectively annealed at higher temperatures in both hardware and software.  We generate similar constraint satisfaction instances with limited range of coupling strength and perform a similar comparison to classical algorithms.  On these instances the D-Wave Vesuvius processor, run with a fixed 20\si{\micro\second} anneal time, shows a scaling advantage in the median case over the software solvers for the hardest regime studied.  This scaling advantage implies that quantum speedup is not ruled out for these problems.  Our results support the hypothesis that performance of D-Wave Vesuvius processors is strongly influenced by analog control error, which can be reduced and mitigated as the technology matures.
\end{abstract}

\maketitle

\section{Introduction}

Following the recent introduction of D-Wave quantum annealing processors, a wealth of research has aimed to characterize the performance of this new platform, in particular pitting it against classical competition \cite{henfl,jking2015,Venturelli2014,Roennow2014,Martin2015,Boixo2013,Boixo2014,Katzgraber2015,boixo2014evidence,mcGeoch2013,Albash2015}.  D-Wave processors take as input spin glass instances in the Ising model, and it is straightforward to express a variety of NP-hard problems in this format \cite{Lucas2014}.  However, the energy landscape of some instances may be more amenable to solution by thermal or combinatorial methods than quantum methods \cite{Katzgraber2014,Steiger2015}, and input to current D-Wave processors must be reasonably robust to analog control error if we are to observe the mechanics of the underlying quantum annealing platform rather than classical noise \cite{Venturelli2014,Zhu2015}.  The selection of appropriate testbeds to use when probing for quantum speedup has recently been the subject of much research.  This research has identified several desirable properties of input sets, including the existence of a nonzero-temperature spin glass phase transition \cite{Katzgraber2014}, foreknowledge of the ground state energy and possibly ground states \cite{henfl}, tunable difficulty, and robustness to analog control error and thermal effects \cite{Zhu2015}.

Randomly generated instances of constraint satisfaction problems (CSPs) or {\em satisfiability} problems are an attractive target: they are well-understood from a statistical physics perspective, and their difficulty can be tuned by a single parameter: the constraint-to-variable ratio $\alpha$ \cite{Mezard2009}.  However, direct solution of these instances requires, in general, the ability to couple arbitrary pairs of qubits in the processor.  While this can be done indirectly in a D-Wave processor through creation of logical qubits \cite{Venturelli2014,Klymko2014,Cai2014,Boothby2015,Vinci2015}, this may amplify control error and obscure the underlying mechanics of the processor \cite{Venturelli2014}.

Hen \cite{henaqc2014} managed this issue by constructing constraint satisfaction problems that can be directly embedded in an arbitrary qubit connectivity graph; each constraint is a {\em frustrated loop}, i.e.\ a cycle of couplers of which an odd number are antiferromagnetic.  These problems have two desirable properties: a {\em planted} (foreknown) ground state and difficulty that can be tuned with the parameter $\alpha$.  Hen et al.\ \cite{henfl} show that performance scaling for the D-Wave processor is superior to the best of a suite of classical solvers in one region of $\alpha$, but is worse in the region of $\alpha$ encompassing the hardest instances.  However, their instances are constructed in such a way that for a fixed value of $\alpha$, thermal effects and analog error are increasingly amplified by normalization as the problems increase in size.

Here we present a simple modification of the construction of these instances that curtails this effect, putting the analog and digital solvers on more level ground and reducing unwanted thermal behavior.  On these {\em range-limited} instances, we find that a D-Wave Vesuvius processor shows better performance scaling than classical competition for all values of $\alpha$ tested.  This competition consists of the two best-performing classical software solvers studied by Hen et al.: the zero-temperature {\em Hamze-de~Freitas-Selby} (HFS) algorithm \cite{Hamze2004,Selby2014,selbycode,selbyblog}, and a solver version of simulated annealing (SAS) \cite{Kirkpatrick1983}.  Hen et al.\ also showed very strong correlation between success probabilities in the D-Wave processor and a thermal Gibbs state approximated using standard simulated annealing (SAA) \cite{henfl}.  Moderating the coupling range of the input instances, and therefore the temperature relative to the final gap of the time-dependent Hamiltonian, reduces correlation with the thermal model.

\section{Quantum annealing and the D-Wave platform\label{sec:qa}}

Quantum annealing in the Ising model aims to find low-energy states in a system of $n$ interacting spins via evolution of the time-dependent Hamiltonian
\begin{equation}
\label{eqn:qahamiltonian}
  \fH_S(t)= \frac 12 \sum_iA(t)\sigma_i^x + B(t)\fH_I
\end{equation}
where $0 \leq t \leq t_f$, $t_f$ is the run time of the QA algorithm, $A(0)\gg B(0)$, $A(t_f)\ll B(t_f)$ and $\fH_I$ is the time-independent Ising problem Hamiltonian:
\begin{equation}
\label{eqn:probleminstance}
\fH_I =  \sum_{i<j}J_{ij}\sigma^z_i\sigma^z_j + \sum_ih_i\sigma^z_i.
\end{equation}
We refer to the Ising Hamiltonian as $\fH_I=(h,J)$, where the biases $h$ and pairwise couplings $J$ encode the optimization problem (i.e.\ energy function) we wish to solve (i.e.\ minimize).  

In a D-Wave quantum annealing processor \cite{Johnson2011, bunyk2014architectural}, not all pairs of qubits are coupled, and therefore the set of nonzero entries of $J$ must adhere to the physical constraints of the processor.  One can view $(h,J)$ as a set of vertex and edge weights, respectively, of the {\em qubit connectivity graph}, whose vertices correspond to qubits and whose edges correspond to couplers.

\section{Frustrated loop problems and limited coupling range\label{sec:fl}}

For a particular hardware qubit connectivity graph $G$ with $n$ vertices (qubits), and a particular constraint-to-qubit ratio $\alpha$, Hen et al.\ construct a {\em frustrated loop} instance using $k=\textit{roundoff}(\alpha n)$ loops like so:
\begin{enumerate}
\item For $i=1,\ldots,k$, loop $\ell_i$ is a cycle in $G$ chosen as the first cycle generated by the edges of a random walk in $G$ starting at a random vertex.  If $\ell_i$ contains fewer than 8 vertices, it is discarded and generated anew.
\item The constraint Ising Hamiltonian $J_i$ corresponding to $\ell_i$ has value $-1$ on every edge of $\ell_i$ except for a randomly selected edge $e_i$ of $\ell_i$, where $J_i$ has value $+1$.  $J_i$ is zero elsewhere.
\item The final Ising Hamiltonian is $(h,J)$, where $h$ is the zero vector and $J=\sum_{i=1}^kJ_i$.
\end{enumerate}

\noindent Any instance constructed with this method has integer-valued $h$ and $J$, but the {\em coupling range} $R=\max_{i,j}\{|J_{ij}|\}$, i.e.\ the maximum magnitude of any entry in $J$, is not necessarily bounded.  Moreover, typical instances constructed at a fixed ratio $\alpha$ on increasingly large subgrids of the D-Wave processor have increasing range limits $R$ \cite{henfl,supp}.  Since input $(h,J)$ to the D-Wave processor must be normalized to within the range $[-1,1]$, coupling range $R$ necessitates scaling by a factor of $1/R$.  This scale factor creates two complications when studying the efficacy of the quantum annealing algorithm on practical hardware.  First, the operating temperature of the processor relative to the magnitude of the input increases with $R$, thus increasing undesirable thermal effects.  Second, each coupler and local field is subject to analog control error on the order of $\delta J \sim 0.035$ and $\delta h \sim 0.05$ respectively. The magnitude of the errors $\delta h$ and $\delta J$ are relative to normalized full energy scale $J =1$. Note that errors are present even for $h = 0$ or $J= 0$. For scale factors of $R$, analog control error relative to the magnitude of the desired input is increased by a factor of $R$. Thus, the deviation of the actual input from the desired input increases with increasing $R$. In the range-unlimited instances studied by Hen et al.\ this amplification factor is as high as 17, and can be dictated by a single coupler that happens to be in disproportionately many loops.  Since $R$ grows with instance size $n$, the D-Wave processor is penalized on larger instances.

In order to address this issue, we construct each instance with respect to an integer coupling range $R\geq 2$, so that in our instances each entry of $h$ and $J$ is an integer between $-R$ and $R$.  To do this, when selecting a candidate for $\ell_i$ via random walk we ignore edges of $G$ on which $|\sum_{j=1}^{i-1}J_i|$ is already $R$.  This ensures that the final Hamiltonian $(h,J)$ has all entries in the range $[-R,R]$, so when the instance is necessarily normalized to the range $[-1,1]$ as input to the D-Wave processor, it is scaled down by no more than a factor of $1/R$.  Consequently, analog control errors and thermal effects in our instances are relatively amplified by no more than a factor of $R$ where $R$ is independent of instance size $n$.

There is another, less crucial modification of the construction: While Hen et al.\ reject and resample a choice of $\ell_i$ if it is too short, we reject the choice if it is contained in a single eight-qubit {\em unit cell} \cite{bunyk2014architectural}.  Thus we sometimes allow loops of length 6, and sometimes forbid loops of length 8.  This modification should in principle allow for greater frustration and less domain clustering within unit cells.

In any meaningful study of analog quantum annealing processors it is desirable to limit relative amplification of analog control error and unwanted thermal effects if it can be done without otherwise materially detracting from the experiment.  In this work we consider $R\in\{2,3,\infty\}$ and $\alpha \in [0.1,0.5]$; these values of $\alpha$ include the hardest regime.  Implications of our choice of coupling range and loop selection criteria are considered in greater detail in the Supplemental Material \cite{supp}.

All of these frustrated loop instances will, by construction, have $\uparrow\uparrow\cdots\uparrow$ and $\downarrow\downarrow\cdots\downarrow$ as planted ground states.  Hen et al.\ describe these instances as being constructed with respect to an arbitrary antipodal pair of planted solutions, but our construction is equivalent under change of variables (Ising spin reversal) both in theory and, due to the application of random spin reversals in hardware, in practice.

\section{Experimental results}

We compare performance of a D-Wave quantum annealing processor to HFS and SAS, both described by Hen et al.\ \cite{henfl}.  Our instances are constructed on subgraphs of a {\em Chimera graph} $C_L$ \cite{bunyk2014architectural}, in which $n$ of $N=8L^2$ qubits are functional, for $4\leq L\leq 8$ (see Supplemental Material \cite{supp}).  Following their methodology, we run SAS on a linear schedule of optimal length in inverse temperature spanning the range $\beta\in[0.01,5]$ after scaling the input to the range $[-1,1]$.  Scaling of the Hamiltonian has no bearing on HFS, which is a large-neighborhood zero-temperature search that exploits low-treewidth induced subgraphs in the Chimera architecture.

In order to remain consistent with previous probes for quantum speedup \cite{henfl,Roennow2014}, we assume that classical algorithms are run by a perfectly parallel oracle that allows all $n$ sites to be updated simultaneously in simulated annealing, and allows all possible cell updates in HFS to be performed in parallel.  Going even further, we simply divide SAS running time by $n$ and divide HFS running time by $L = \sqrt{N/8}$, the maximum possible number of parallel cell updates at any point in the algorithm.  Further detail on experimental methods, benchmarking and data analysis is given in the Supplemental Material \cite{supp}.  To account for differences between implementations and hardware, we use the assumption of Hen et al.\ that each Monte Carlo sweep takes time $\tau_{\textrm{SA}}=3.54\si{\micro\second}$.  We assume that each HFS unit cell update takes $1\si{\micro\second}$.

The D-Wave processor used was a D-Wave Two V6 processor of the same architecture and fabrication lot as the processor used by Hen et al.\ \cite{henfl}

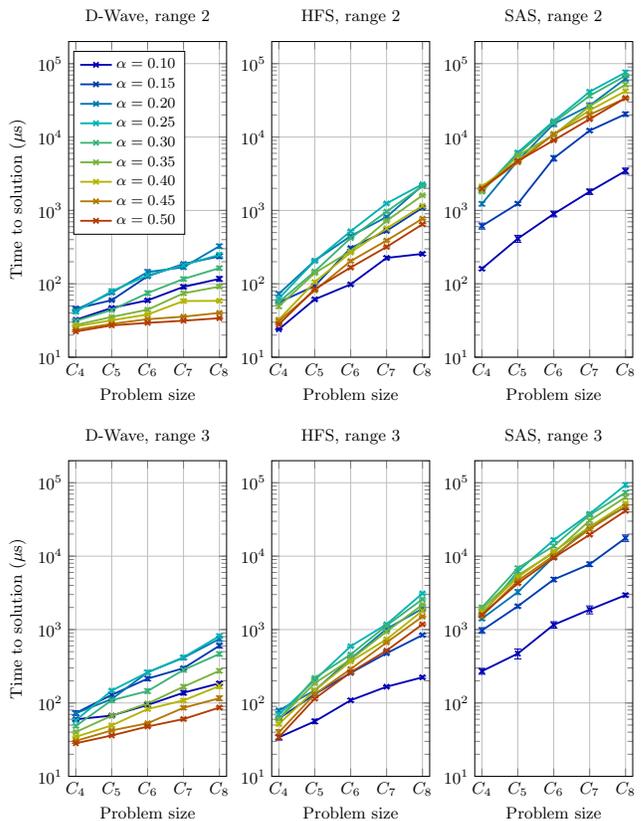
\begin{figure}
\setlength{\figurewidth}{3cm}%
\setlength{\figureheight}{6cm}%
\scalebox{.70}{%
%
%
%
\definecolor{mycolor1}{rgb}{0.00000,0.23333,0.70000}%
\definecolor{mycolor2}{rgb}{0.00000,0.46667,0.70000}%
\definecolor{mycolor3}{rgb}{0.00000,0.70000,0.70000}%
\definecolor{mycolor4}{rgb}{0.23333,0.70000,0.46667}%
\definecolor{mycolor5}{rgb}{0.46667,0.70000,0.23333}%
\definecolor{mycolor6}{rgb}{0.70000,0.70000,0.00000}%
\definecolor{mycolor7}{rgb}{0.70000,0.46667,0.00000}%
\definecolor{mycolor8}{rgb}{0.70000,0.23333,0.00000}%
\begin{tikzpicture}

\begin{axis}[%
width=\figurewidth,
height=\figureheight,
scale only axis,
xmin=3.8,
xmax=8.2,
xlabel={Problem size},
xmajorgrids,
ymode=log,
ymin=10,
ymax=200000,
yminorticks=true,
ylabel={Time to solution ($\mu$s)},
ymajorgrids,
title={D-Wave, range 2},
legend style={at={(0.03,0.97)},anchor=north west,draw=black,fill=white,legend cell align=left},
legend style={font=\footnotesize},xtick={4,5,6,7,8},xticklabels={$C_4$,$C_5$,$C_6$,$C_7$,$C_8$}
]
\addplot [color=black!30!blue,solid,line width=1.0pt,mark size=2.0pt,mark=x,mark options={solid}]
 plot [error bars/.cd, y dir = both, y explicit]
 table[row sep=crcr, y error plus index=2, y error minus index=3]{
4	32.3276716166663	0.302696043081561	0.302696043081561	\\
5	46.9968501779953	1.17130413514698	1.17130413514698	\\
6	59.5308253904455	2.58824768514926	2.58824768514926	\\
7	90.9327982740047	4.92199129972623	4.92199129972623	\\
8	116.76845017638	8.43522109016723	8.43522109016723	\\
};
\addlegendentry{$\alpha = 0.10$};

\addplot [color=mycolor1,solid,line width=1.0pt,mark size=2.0pt,mark=x,mark options={solid}]
 plot [error bars/.cd, y dir = both, y explicit]
 table[row sep=crcr, y error plus index=2, y error minus index=3]{
4	45.8654502360608	3.06234456695314	3.06234456695314	\\
5	59.8185770667225	1.33290507566914	1.33290507566914	\\
6	126.937221298145	5.33590729205108	5.33590729205106	\\
7	186.423652330568	7.28938562591006	7.28938562591006	\\
8	237.257878570049	15.6721028372917	15.6721028372917	\\
};
\addlegendentry{$\alpha = 0.15$};

\addplot [color=mycolor2,solid,line width=1.0pt,mark size=2.0pt,mark=x,mark options={solid}]
 plot [error bars/.cd, y dir = both, y explicit]
 table[row sep=crcr, y error plus index=2, y error minus index=3]{
4	44.1474723709369	0.960035907985507	0.960035907985507	\\
5	75.8588908839113	3.41309489018673	3.41309489018673	\\
6	144.66211818489	11.3836589979547	11.3836589979547	\\
7	168.302217089816	9.041160703039	9.041160703039	\\
8	325.864837268955	20.6884891851791	20.6884891851791	\\
};
\addlegendentry{$\alpha = 0.20$};

\addplot [color=mycolor3,solid,line width=1.0pt,mark size=2.0pt,mark=x,mark options={solid}]
 plot [error bars/.cd, y dir = both, y explicit]
 table[row sep=crcr, y error plus index=2, y error minus index=3]{
4	41.6650734090452	2.12559852525717	2.12559852525717	\\
5	79.3259661366667	5.90494185604631	5.90494185604631	\\
6	130.141186608464	6.95330982203174	6.95330982203176	\\
7	180.692722751406	9.50712825707268	9.50712825707268	\\
8	248.555036376348	15.0033267739111	15.0033267739111	\\
};
\addlegendentry{$\alpha = 0.25$};

\addplot [color=mycolor4,solid,line width=1.0pt,mark size=2.0pt,mark=x,mark options={solid}]
 plot [error bars/.cd, y dir = both, y explicit]
 table[row sep=crcr, y error plus index=2, y error minus index=3]{
4	31.3486262679155	0.79917290172358	0.79917290172358	\\
5	43.956173480299	1.14531030320619	1.14531030320619	\\
6	74.8406530938881	5.16041621699797	5.16041621699797	\\
7	116.459670368573	6.54517208679785	6.54517208679785	\\
8	163.759823756835	8.37489689154472	8.37489689154472	\\
};
\addlegendentry{$\alpha = 0.30$};

\addplot [color=mycolor5,solid,line width=1.0pt,mark size=2.0pt,mark=x,mark options={solid}]
 plot [error bars/.cd, y dir = both, y explicit]
 table[row sep=crcr, y error plus index=2, y error minus index=3]{
4	27.4824628316128	0.292213383953811	0.292213383953811	\\
5	35.3682688233578	0.407039118405322	0.407039118405322	\\
6	44.4081686709836	1.67760606658896	1.67760606658896	\\
7	74.2725727055745	3.63478553669243	3.63478553669243	\\
8	91.5611588422833	3.98331357298306	3.98331357298306	\\
};
\addlegendentry{$\alpha = 0.35$};

\addplot [color=mycolor6,solid,line width=1.0pt,mark size=2.0pt,mark=x,mark options={solid}]
 plot [error bars/.cd, y dir = both, y explicit]
 table[row sep=crcr, y error plus index=2, y error minus index=3]{
4	26.267512870847	0.248974656107997	0.248974656107997	\\
5	31.8781536323523	0.30717211395088	0.307172113950877	\\
6	38.1678124935977	0.710112738251517	0.710112738251517	\\
7	58.306816159289	2.84604718535383	2.84604718535383	\\
8	58.6234993688315	2.35425559982789	2.35425559982789	\\
};
\addlegendentry{$\alpha = 0.40$};

\addplot [color=mycolor7,solid,line width=1.0pt,mark size=2.0pt,mark=x,mark options={solid}]
 plot [error bars/.cd, y dir = both, y explicit]
 table[row sep=crcr, y error plus index=2, y error minus index=3]{
4	23.6551289388046	0.128257165960274	0.128257165960274	\\
5	28.4969422927821	0.367737509332208	0.367737509332208	\\
6	32.9757116472971	0.37106564491306	0.37106564491306	\\
7	35.7155198572372	0.552155461288187	0.552155461288187	\\
8	40.1477232620536	0.820140228374001	0.820140228374001	\\
};
\addlegendentry{$\alpha = 0.45$};

\addplot [color=mycolor8,solid,line width=1.0pt,mark size=2.0pt,mark=x,mark options={solid}]
 plot [error bars/.cd, y dir = both, y explicit]
 table[row sep=crcr, y error plus index=2, y error minus index=3]{
4	22.4201701178993	0.152689437985078	0.152689437985078	\\
5	27.1249767872441	0.233855168095396	0.233855168095396	\\
6	29.3942585502291	0.308360087610474	0.308360087610474	\\
7	31.4405475679521	0.23182556800047	0.23182556800047	\\
8	34.018955119435	0.361484438202112	0.361484438202112	\\
};
\addlegendentry{$\alpha = 0.50$};

\end{axis}
\end{tikzpicture}%
%
%
%
\definecolor{mycolor1}{rgb}{0.00000,0.23333,0.70000}%
\definecolor{mycolor2}{rgb}{0.00000,0.46667,0.70000}%
\definecolor{mycolor3}{rgb}{0.00000,0.70000,0.70000}%
\definecolor{mycolor4}{rgb}{0.23333,0.70000,0.46667}%
\definecolor{mycolor5}{rgb}{0.46667,0.70000,0.23333}%
\definecolor{mycolor6}{rgb}{0.70000,0.70000,0.00000}%
\definecolor{mycolor7}{rgb}{0.70000,0.46667,0.00000}%
\definecolor{mycolor8}{rgb}{0.70000,0.23333,0.00000}%
\begin{tikzpicture}

\begin{axis}[%
width=\figurewidth,
height=\figureheight,
scale only axis,
xmin=3.8,
xmax=8.2,
xlabel={Problem size},
xmajorgrids,
ymode=log,
ymin=10,
ymax=200000,
yminorticks=true,
ymajorgrids,
title={HFS, range 2},
legend style={font=\footnotesize},xtick={4,5,6,7,8},xticklabels={$C_4$,$C_5$,$C_6$,$C_7$,$C_8$}
]
\addplot [color=black!30!blue,solid,line width=1.0pt,mark size=2.0pt,mark=x,mark options={solid},forget plot]
 plot [error bars/.cd, y dir = both, y explicit]
 table[row sep=crcr, y error plus index=2, y error minus index=3]{
4	24.022	0.274939250729304	0.274939250729304	\\
5	61.45	2.23102610416764	2.23102610416764	\\
6	98.286	3.52002764116912	3.52002764116912	\\
7	225.414	10.8015173367808	10.8015173367808	\\
8	255.908	10.6625149577961	10.6625149577961	\\
};
\addplot [color=mycolor1,solid,line width=1.0pt,mark size=2.0pt,mark=x,mark options={solid},forget plot]
 plot [error bars/.cd, y dir = both, y explicit]
 table[row sep=crcr, y error plus index=2, y error minus index=3]{
4	55.944	1.1018521126168	1.1018521126168	\\
5	93.525	4.02530471866143	4.02530471866143	\\
6	307.005	11.6714694104126	11.6714694104126	\\
7	526.2845	17.6872453592524	17.6872453592525	\\
8	1087.248	29.8187131119312	29.8187131119312	\\
};
\addplot [color=mycolor2,solid,line width=1.0pt,mark size=2.0pt,mark=x,mark options={solid},forget plot]
 plot [error bars/.cd, y dir = both, y explicit]
 table[row sep=crcr, y error plus index=2, y error minus index=3]{
4	73.196	3.45680973348743	3.45680973348743	\\
5	206.4375	6.59134201354621	6.59134201354621	\\
6	447.609	12.5575399983759	12.5575399983759	\\
7	815.416	21.115131844078	21.115131844078	\\
8	2183.972	51.2407731482735	51.2407731482735	\\
};
\addplot [color=mycolor3,solid,line width=1.0pt,mark size=2.0pt,mark=x,mark options={solid},forget plot]
 plot [error bars/.cd, y dir = both, y explicit]
 table[row sep=crcr, y error plus index=2, y error minus index=3]{
4	63.194	1.94271411642016	1.94271411642017	\\
5	205.43	6.10859353001285	6.10859353001285	\\
6	520.461	13.7643720138965	13.7643720138965	\\
7	1245.762	30.4204003385353	30.4204003385353	\\
8	2275.06	42.4728659775196	42.4728659775196	\\
};
\addplot [color=mycolor4,solid,line width=1.0pt,mark size=2.0pt,mark=x,mark options={solid},forget plot]
 plot [error bars/.cd, y dir = both, y explicit]
 table[row sep=crcr, y error plus index=2, y error minus index=3]{
4	56.442	1.39232703578382	1.39232703578382	\\
5	147.3025	5.27447284700699	5.27447284700699	\\
6	421.062	11.2924610105891	11.2924610105891	\\
7	963.466	24.1899358617793	24.1899358617793	\\
8	2138.44	39.8563787985177	39.8563787985177	\\
};
\addplot [color=mycolor5,solid,line width=1.0pt,mark size=2.0pt,mark=x,mark options={solid},forget plot]
 plot [error bars/.cd, y dir = both, y explicit]
 table[row sep=crcr, y error plus index=2, y error minus index=3]{
4	49.406	2.99133951802736	2.99133951802736	\\
5	140.275	4.93444259503786	4.93444259503786	\\
6	279.717	7.72259992162935	7.72259992162935	\\
7	720.706	16.6629444492178	16.6629444492178	\\
8	1607.02	30.5447308153236	30.5447308153236	\\
};
\addplot [color=mycolor6,solid,line width=1.0pt,mark size=2.0pt,mark=x,mark options={solid},forget plot]
 plot [error bars/.cd, y dir = both, y explicit]
 table[row sep=crcr, y error plus index=2, y error minus index=3]{
4	32.622	1.38384646223165	1.38384646223164	\\
5	105.645	4.02649819169061	4.02649819169061	\\
6	267.612	7.74386359793505	7.74386359793505	\\
7	569.107	15.0214911442305	15.0214911442305	\\
8	1153.824	28.338597355905	28.338597355905	\\
};
\addplot [color=mycolor7,solid,line width=1.0pt,mark size=2.0pt,mark=x,mark options={solid},forget plot]
 plot [error bars/.cd, y dir = both, y explicit]
 table[row sep=crcr, y error plus index=2, y error minus index=3]{
4	31.156	1.55501835425892	1.55501835425892	\\
5	81.3925	2.4896570330353	2.4896570330353	\\
6	204.081	5.86812353941818	5.86812353941818	\\
7	388.619	10.5732271389643	10.5732271389643	\\
8	771.2	18.7283408435118	18.7283408435118	\\
};
\addplot [color=mycolor8,solid,line width=1.0pt,mark size=2.0pt,mark=x,mark options={solid},forget plot]
 plot [error bars/.cd, y dir = both, y explicit]
 table[row sep=crcr, y error plus index=2, y error minus index=3]{
4	28	0	0	\\
5	84.7625	3.24389682582792	3.24389682582792	\\
6	166.95	5.80579083391808	5.80579083391808	\\
7	318.1185	8.86650001819049	8.86650001819049	\\
8	647.476	18.1894551947076	18.1894551947076	\\
};
\end{axis}
\end{tikzpicture}%
%
%
%
\definecolor{mycolor1}{rgb}{0.00000,0.23333,0.70000}%
\definecolor{mycolor2}{rgb}{0.00000,0.46667,0.70000}%
\definecolor{mycolor3}{rgb}{0.00000,0.70000,0.70000}%
\definecolor{mycolor4}{rgb}{0.23333,0.70000,0.46667}%
\definecolor{mycolor5}{rgb}{0.46667,0.70000,0.23333}%
\definecolor{mycolor6}{rgb}{0.70000,0.70000,0.00000}%
\definecolor{mycolor7}{rgb}{0.70000,0.46667,0.00000}%
\definecolor{mycolor8}{rgb}{0.70000,0.23333,0.00000}%
\begin{tikzpicture}

\begin{axis}[%
width=\figurewidth,
height=\figureheight,
scale only axis,
xmin=3.8,
xmax=8.2,
xlabel={Problem size},
xmajorgrids,
ymode=log,
ymin=10,
ymax=200000,
yminorticks=true,
ymajorgrids,
title={SAS, range 2},
legend style={font=\footnotesize},xtick={4,5,6,7,8},xticklabels={$C_4$,$C_5$,$C_6$,$C_7$,$C_8$}
]
\addplot [color=black!30!blue,solid,line width=1.0pt,mark size=2.0pt,mark=x,mark options={solid},forget plot]
 plot [error bars/.cd, y dir = both, y explicit]
 table[row sep=crcr, y error plus index=2, y error minus index=3]{
4	160.588983717136	6.75944057246269	6.75944057246269	\\
5	413.01495858617	45.2254628410316	45.2254628410316	\\
6	896.407057430769	79.4205834276432	79.4205834276432	\\
7	1803.24982434889	152.294876345667	152.294876345667	\\
8	3455.17722415048	320.2235932764	320.2235932764	\\
};
\addplot [color=mycolor1,solid,line width=1.0pt,mark size=2.0pt,mark=x,mark options={solid},forget plot]
 plot [error bars/.cd, y dir = both, y explicit]
 table[row sep=crcr, y error plus index=2, y error minus index=3]{
4	613.940121313706	58.0434010774346	58.0434010774346	\\
5	1236.77370373542	59.6977613088493	59.6977613088493	\\
6	5146.28835700175	440.68437108926	440.68437108926	\\
7	12240.1476828371	541.431339890049	541.431339890049	\\
8	20530.5069578641	1329.92097438292	1329.92097438292	\\
};
\addplot [color=mycolor2,solid,line width=1.0pt,mark size=2.0pt,mark=x,mark options={solid},forget plot]
 plot [error bars/.cd, y dir = both, y explicit]
 table[row sep=crcr, y error plus index=2, y error minus index=3]{
4	1230.38327714987	74.3454129325078	74.3454129325078	\\
5	4757.52728464854	250.034309824392	250.034309824392	\\
6	15125.2014667414	1232.84860016613	1232.84860016613	\\
7	26815.341708466	2385.04478222985	2385.04478222985	\\
8	62392.1413837385	4221.63125797168	4221.63125797168	\\
};
\addplot [color=mycolor3,solid,line width=1.0pt,mark size=2.0pt,mark=x,mark options={solid},forget plot]
 plot [error bars/.cd, y dir = both, y explicit]
 table[row sep=crcr, y error plus index=2, y error minus index=3]{
4	1810.39093220111	63.314767905252	63.314767905252	\\
5	6149.48174606042	341.600964837587	341.600964837587	\\
6	16506.3392594403	572.787207576621	572.787207576621	\\
7	41362.4547345749	1598.83057112167	1598.83057112167	\\
8	75877.7258528724	5315.44617993465	5315.44617993465	\\
};
\addplot [color=mycolor4,solid,line width=1.0pt,mark size=2.0pt,mark=x,mark options={solid},forget plot]
 plot [error bars/.cd, y dir = both, y explicit]
 table[row sep=crcr, y error plus index=2, y error minus index=3]{
4	1870.06681191714	73.9861201854505	73.9861201854505	\\
5	5813.36626818437	355.861212601258	355.861212601258	\\
6	15715.788547919	591.857693623489	591.857693623489	\\
7	36470.374375328	1843.88845481741	1843.88845481741	\\
8	68302.5381790815	2868.06826054437	2868.06826054437	\\
};
\addplot [color=mycolor5,solid,line width=1.0pt,mark size=2.0pt,mark=x,mark options={solid},forget plot]
 plot [error bars/.cd, y dir = both, y explicit]
 table[row sep=crcr, y error plus index=2, y error minus index=3]{
4	1820.3767876469	69.5818389966757	69.5818389966757	\\
5	5585.81082930501	229.094647918036	229.094647918036	\\
6	10638.9178109994	546.993889864922	546.993889864922	\\
7	26410.27569652	902.874163221157	902.874163221157	\\
8	51820.3345475642	2786.81204165552	2786.81204165552	\\
};
\addplot [color=mycolor6,solid,line width=1.0pt,mark size=2.0pt,mark=x,mark options={solid},forget plot]
 plot [error bars/.cd, y dir = both, y explicit]
 table[row sep=crcr, y error plus index=2, y error minus index=3]{
4	2127.15563165598	73.9522526926544	73.9522526926544	\\
5	4806.30210653563	98.3308281035779	98.3308281035779	\\
6	10971.0619732818	308.534591455878	308.534591455878	\\
7	23136.4624236989	933.496670522516	933.496670522516	\\
8	42248.7892797674	1776.71149835858	1776.71149835858	\\
};
\addplot [color=mycolor7,solid,line width=1.0pt,mark size=2.0pt,mark=x,mark options={solid},forget plot]
 plot [error bars/.cd, y dir = both, y explicit]
 table[row sep=crcr, y error plus index=2, y error minus index=3]{
4	1984.62391471225	48.9657522174266	48.9657522174266	\\
5	4518.64633529082	139.018369541893	139.018369541893	\\
6	10944.8938566404	356.902058182088	356.902058182088	\\
7	20142.8565130043	690.502560471643	690.502560471643	\\
8	33761.9896940394	1056.56405761177	1056.56405761177	\\
};
\addplot [color=mycolor8,solid,line width=1.0pt,mark size=2.0pt,mark=x,mark options={solid},forget plot]
 plot [error bars/.cd, y dir = both, y explicit]
 table[row sep=crcr, y error plus index=2, y error minus index=3]{
4	1986.7822287613	39.9230917680102	39.9230917680102	\\
5	4720.56481082973	121.55218692405	121.55218692405	\\
6	9028.12439035051	171.605902097996	171.605902097996	\\
7	17521.722563938	374.980623469917	374.980623469917	\\
8	33959.198108504	731.048769407796	731.048769407796	\\
};
\end{axis}
\end{tikzpicture}%
}\vspace{.2cm}
\scalebox{.70}{%
%
%
%
\definecolor{mycolor1}{rgb}{0.00000,0.23333,0.70000}%
\definecolor{mycolor2}{rgb}{0.00000,0.46667,0.70000}%
\definecolor{mycolor3}{rgb}{0.00000,0.70000,0.70000}%
\definecolor{mycolor4}{rgb}{0.23333,0.70000,0.46667}%
\definecolor{mycolor5}{rgb}{0.46667,0.70000,0.23333}%
\definecolor{mycolor6}{rgb}{0.70000,0.70000,0.00000}%
\definecolor{mycolor7}{rgb}{0.70000,0.46667,0.00000}%
\definecolor{mycolor8}{rgb}{0.70000,0.23333,0.00000}%
\begin{tikzpicture}

\begin{axis}[%
width=\figurewidth,
height=\figureheight,
scale only axis,
xmin=3.8,
xmax=8.2,
xlabel={Problem size},
xmajorgrids,
ymode=log,
ymin=10,
ymax=200000,
yminorticks=true,
ylabel={Time to solution ($\mu$s)},
ymajorgrids,
title={D-Wave, range 3},
legend style={font=\footnotesize},xtick={4,5,6,7,8},xticklabels={$C_4$,$C_5$,$C_6$,$C_7$,$C_8$}
]
\addplot [color=black!30!blue,solid,line width=1.0pt,mark size=2.0pt,mark=x,mark options={solid},forget plot]
 plot [error bars/.cd, y dir = both, y explicit]
 table[row sep=crcr, y error plus index=2, y error minus index=3]{
4	59.8196749785874	2.07383328626239	2.07383328626239	\\
5	67.2255671256053	1.91223721821017	1.91223721821017	\\
6	93.9194245964572	5.8265024300365	5.8265024300365	\\
7	137.935585082976	9.63178789454625	9.63178789454625	\\
8	185.301741408938	8.93734719777402	8.93734719777402	\\
};
\addplot [color=mycolor1,solid,line width=1.0pt,mark size=2.0pt,mark=x,mark options={solid},forget plot]
 plot [error bars/.cd, y dir = both, y explicit]
 table[row sep=crcr, y error plus index=2, y error minus index=3]{
4	72.9842915108241	5.41193689516085	5.41193689516085	\\
5	128.784070691254	5.33583033795159	5.33583033795159	\\
6	214.598559857621	13.9528706868386	13.9528706868386	\\
7	297.60731355053	13.9417925943346	13.9417925943346	\\
8	602.681961863061	41.3768072819151	41.3768072819151	\\
};
\addplot [color=mycolor2,solid,line width=1.0pt,mark size=2.0pt,mark=x,mark options={solid},forget plot]
 plot [error bars/.cd, y dir = both, y explicit]
 table[row sep=crcr, y error plus index=2, y error minus index=3]{
4	72.2308341351822	3.67049343274485	3.67049343274485	\\
5	113.130693748095	5.51764498721535	5.51764498721535	\\
6	261.565779431999	12.8720233532894	12.8720233532894	\\
7	413.099152065278	21.3573962664819	21.3573962664819	\\
8	746.04539400412	55.642986065257	55.642986065257	\\
};
\addplot [color=mycolor3,solid,line width=1.0pt,mark size=2.0pt,mark=x,mark options={solid},forget plot]
 plot [error bars/.cd, y dir = both, y explicit]
 table[row sep=crcr, y error plus index=2, y error minus index=3]{
4	57.9473520871594	2.15371600090432	2.15371600090432	\\
5	147.517908442241	5.21756359723108	5.21756359723108	\\
6	260.871320238428	13.281122699638	13.2811226996381	\\
7	419.748655367071	22.673052970327	22.673052970327	\\
8	815.52077390965	51.3298978433318	51.3298978433318	\\
};
\addplot [color=mycolor4,solid,line width=1.0pt,mark size=2.0pt,mark=x,mark options={solid},forget plot]
 plot [error bars/.cd, y dir = both, y explicit]
 table[row sep=crcr, y error plus index=2, y error minus index=3]{
4	48.5506138901175	1.90129421026781	1.90129421026781	\\
5	108.902086203446	7.62338157691045	7.62338157691045	\\
6	145.277720673085	9.92257024343954	9.92257024343954	\\
7	282.161639266469	17.551556260627	17.551556260627	\\
8	464.983456689248	28.8167000904289	28.8167000904289	\\
};
\addplot [color=mycolor5,solid,line width=1.0pt,mark size=2.0pt,mark=x,mark options={solid},forget plot]
 plot [error bars/.cd, y dir = both, y explicit]
 table[row sep=crcr, y error plus index=2, y error minus index=3]{
4	40.1379005273088	1.01830557503629	1.01830557503629	\\
5	66.9613349548124	3.35042168750877	3.35042168750876	\\
6	98.1984610394277	4.79687178097601	4.79687178097601	\\
7	166.738730148324	9.98136581957991	9.98136581957991	\\
8	274.099467992011	15.8223484250389	15.8223484250389	\\
};
\addplot [color=mycolor6,solid,line width=1.0pt,mark size=2.0pt,mark=x,mark options={solid},forget plot]
 plot [error bars/.cd, y dir = both, y explicit]
 table[row sep=crcr, y error plus index=2, y error minus index=3]{
4	34.5212121243389	0.649721696088676	0.649721696088676	\\
5	49.4352502237016	1.87560190314767	1.87560190314767	\\
6	82.5819464229003	4.04308978793249	4.04308978793249	\\
7	108.232840206293	5.46088994070234	5.46088994070234	\\
8	170.557239355907	13.1940935823364	13.1940935823364	\\
};
\addplot [color=mycolor7,solid,line width=1.0pt,mark size=2.0pt,mark=x,mark options={solid},forget plot]
 plot [error bars/.cd, y dir = both, y explicit]
 table[row sep=crcr, y error plus index=2, y error minus index=3]{
4	30.4454511160626	0.259040483526707	0.259040483526707	\\
5	42.2254023713196	0.931913889660386	0.931913889660386	\\
6	52.8992674219683	1.26179725652177	1.26179725652177	\\
7	85.9548999934169	4.6091463519851	4.6091463519851	\\
8	115.969428999632	8.45447198492933	8.45447198492933	\\
};
\addplot [color=mycolor8,solid,line width=1.0pt,mark size=2.0pt,mark=x,mark options={solid},forget plot]
 plot [error bars/.cd, y dir = both, y explicit]
 table[row sep=crcr, y error plus index=2, y error minus index=3]{
4	28.231393686962	0.168479405523009	0.168479405523009	\\
5	36.0518859240978	0.343401647933838	0.343401647933838	\\
6	47.7750358322172	1.52197578939238	1.52197578939238	\\
7	60.3818843888493	2.24658108998399	2.24658108998399	\\
8	86.5476055728168	3.39657609755362	3.39657609755362	\\
};
\end{axis}
\end{tikzpicture}%
%
%
%
\definecolor{mycolor1}{rgb}{0.00000,0.23333,0.70000}%
\definecolor{mycolor2}{rgb}{0.00000,0.46667,0.70000}%
\definecolor{mycolor3}{rgb}{0.00000,0.70000,0.70000}%
\definecolor{mycolor4}{rgb}{0.23333,0.70000,0.46667}%
\definecolor{mycolor5}{rgb}{0.46667,0.70000,0.23333}%
\definecolor{mycolor6}{rgb}{0.70000,0.70000,0.00000}%
\definecolor{mycolor7}{rgb}{0.70000,0.46667,0.00000}%
\definecolor{mycolor8}{rgb}{0.70000,0.23333,0.00000}%
\begin{tikzpicture}

\begin{axis}[%
width=\figurewidth,
height=\figureheight,
scale only axis,
xmin=3.8,
xmax=8.2,
xlabel={Problem size},
xmajorgrids,
ymode=log,
ymin=10,
ymax=200000,
yminorticks=true,
ymajorgrids,
title={HFS, range 3},
legend style={font=\footnotesize},xtick={4,5,6,7,8},xticklabels={$C_4$,$C_5$,$C_6$,$C_7$,$C_8$}
]
\addplot [color=black!30!blue,solid,line width=1.0pt,mark size=2.0pt,mark=x,mark options={solid},forget plot]
 plot [error bars/.cd, y dir = both, y explicit]
 table[row sep=crcr, y error plus index=2, y error minus index=3]{
4	34.054	2.83286235265836	2.83286235265836	\\
5	56.1325	3.88281727873202	3.88281727873202	\\
6	108.831	5.08687677414298	5.08687677414298	\\
7	166.488	4.93159980905557	4.93159980905557	\\
8	223.6	8.10145873492141	8.10145873492141	\\
};
\addplot [color=mycolor1,solid,line width=1.0pt,mark size=2.0pt,mark=x,mark options={solid},forget plot]
 plot [error bars/.cd, y dir = both, y explicit]
 table[row sep=crcr, y error plus index=2, y error minus index=3]{
4	61.27	2.25836228276161	2.25836228276161	\\
5	131.56	4.66375928309458	4.66375928309456	\\
6	257.037	8.46080470264008	8.46080470264005	\\
7	475.426	15.0221066328158	15.0221066328158	\\
8	841.668	27.3770575350836	27.3770575350836	\\
};
\addplot [color=mycolor2,solid,line width=1.0pt,mark size=2.0pt,mark=x,mark options={solid},forget plot]
 plot [error bars/.cd, y dir = both, y explicit]
 table[row sep=crcr, y error plus index=2, y error minus index=3]{
4	79.292	2.98054015567979	2.98054015567979	\\
5	148.28	5.39636180409735	5.39636180409735	\\
6	399.546	12.0033313694142	12.0033313694142	\\
7	1022.063	26.6845572681392	26.6845572681392	\\
8	1926.704	48.7385661651206	48.7385661651206	\\
};
\addplot [color=mycolor3,solid,line width=1.0pt,mark size=2.0pt,mark=x,mark options={solid},forget plot]
 plot [error bars/.cd, y dir = both, y explicit]
 table[row sep=crcr, y error plus index=2, y error minus index=3]{
4	69.722	3.1471607236715	3.1471607236715	\\
5	205.5475	6.65306373086918	6.65306373086918	\\
6	592.767	14.5775586507555	14.5775586507555	\\
7	1179.563	29.1251576099737	29.1251576099737	\\
8	3110.1	69.965399714124	69.965399714124	\\
};
\addplot [color=mycolor4,solid,line width=1.0pt,mark size=2.0pt,mark=x,mark options={solid},forget plot]
 plot [error bars/.cd, y dir = both, y explicit]
 table[row sep=crcr, y error plus index=2, y error minus index=3]{
4	60.582	1.48441216630674	1.48441216630674	\\
5	219.395	6.78638563817518	6.78638563817518	\\
6	458.133	12.5229443476818	12.5229443476818	\\
7	1132.7855	26.162697232443	26.162697232443	\\
8	2587.452	49.0904223404686	49.0904223404686	\\
};
\addplot [color=mycolor5,solid,line width=1.0pt,mark size=2.0pt,mark=x,mark options={solid},forget plot]
 plot [error bars/.cd, y dir = both, y explicit]
 table[row sep=crcr, y error plus index=2, y error minus index=3]{
4	60.11	1.22368190204701	1.22368190204701	\\
5	185.885	5.88654622222796	5.88654622222796	\\
6	391.878	10.7536392331507	10.7536392331507	\\
7	939.722	20.6169414383966	20.6169414383966	\\
8	2180.408	40.9378729961077	40.9378729961077	\\
};
\addplot [color=mycolor6,solid,line width=1.0pt,mark size=2.0pt,mark=x,mark options={solid},forget plot]
 plot [error bars/.cd, y dir = both, y explicit]
 table[row sep=crcr, y error plus index=2, y error minus index=3]{
4	51.67	1.77376611937602	1.77376611937602	\\
5	148.6475	4.33973813462021	4.33973813462021	\\
6	368.106	10.0616102091067	10.0616102091067	\\
7	736.5295	18.1200086772734	18.1200086772734	\\
8	1799.212	34.9134645374247	34.9134645374247	\\
};
\addplot [color=mycolor7,solid,line width=1.0pt,mark size=2.0pt,mark=x,mark options={solid},forget plot]
 plot [error bars/.cd, y dir = both, y explicit]
 table[row sep=crcr, y error plus index=2, y error minus index=3]{
4	39.744	3.6556569625483	3.6556569625483	\\
5	131.485	4.58020231575563	4.58020231575563	\\
6	289.767	8.05167936150843	8.05167936150843	\\
7	667.898	16.3994218506946	16.3994218506946	\\
8	1500.28	29.9987587330797	29.9987587330797	\\
};
\addplot [color=mycolor8,solid,line width=1.0pt,mark size=2.0pt,mark=x,mark options={solid},forget plot]
 plot [error bars/.cd, y dir = both, y explicit]
 table[row sep=crcr, y error plus index=2, y error minus index=3]{
4	34.108	2.11679399489341	2.11679399489341	\\
5	114.725	3.17363308322606	3.17363308322606	\\
6	261.033	7.09706131291534	7.09706131291534	\\
7	516.8345	13.0536492027137	13.0536492027137	\\
8	1179.212	25.0635839264708	25.0635839264708	\\
};
\end{axis}
\end{tikzpicture}%
%
%
%
\definecolor{mycolor1}{rgb}{0.00000,0.23333,0.70000}%
\definecolor{mycolor2}{rgb}{0.00000,0.46667,0.70000}%
\definecolor{mycolor3}{rgb}{0.00000,0.70000,0.70000}%
\definecolor{mycolor4}{rgb}{0.23333,0.70000,0.46667}%
\definecolor{mycolor5}{rgb}{0.46667,0.70000,0.23333}%
\definecolor{mycolor6}{rgb}{0.70000,0.70000,0.00000}%
\definecolor{mycolor7}{rgb}{0.70000,0.46667,0.00000}%
\definecolor{mycolor8}{rgb}{0.70000,0.23333,0.00000}%
\begin{tikzpicture}

\begin{axis}[%
width=\figurewidth,
height=\figureheight,
scale only axis,
xmin=3.8,
xmax=8.2,
xlabel={Problem size},
xmajorgrids,
ymode=log,
ymin=10,
ymax=200000,
yminorticks=true,
ymajorgrids,
title={SAS, range 3},
legend style={font=\footnotesize},xtick={4,5,6,7,8},xticklabels={$C_4$,$C_5$,$C_6$,$C_7$,$C_8$}
]
\addplot [color=black!30!blue,solid,line width=1.0pt,mark size=2.0pt,mark=x,mark options={solid},forget plot]
 plot [error bars/.cd, y dir = both, y explicit]
 table[row sep=crcr, y error plus index=2, y error minus index=3]{
4	271.128912031822	25.0702766618915	25.0702766618915	\\
5	470.81182387668	72.247894513458	72.247894513458	\\
6	1161.41287944597	117.019815085516	117.019815085516	\\
7	1868.44088609124	230.944081828515	230.944081828515	\\
8	2947.66085570565	144.507514600522	144.507514600522	\\
};
\addplot [color=mycolor1,solid,line width=1.0pt,mark size=2.0pt,mark=x,mark options={solid},forget plot]
 plot [error bars/.cd, y dir = both, y explicit]
 table[row sep=crcr, y error plus index=2, y error minus index=3]{
4	974.107533774184	85.9382772778537	85.9382772778537	\\
5	2077.67682894017	94.3662343125475	94.3662343125475	\\
6	4811.35177271405	307.402441721995	307.402441721995	\\
7	7762.91471903868	535.39570491677	535.395704916771	\\
8	17658.4489460252	1825.96010267267	1825.96010267267	\\
};
\addplot [color=mycolor2,solid,line width=1.0pt,mark size=2.0pt,mark=x,mark options={solid},forget plot]
 plot [error bars/.cd, y dir = both, y explicit]
 table[row sep=crcr, y error plus index=2, y error minus index=3]{
4	1408.6543082317	62.8830527269731	62.8830527269731	\\
5	3243.73193070742	242.91140886114	242.91140886114	\\
6	9743.53602721442	693.595544863612	693.595544863612	\\
7	25347.9268966773	1749.9816712663	1749.9816712663	\\
8	46627.2607694392	3577.49397054995	3577.49397054995	\\
};
\addplot [color=mycolor3,solid,line width=1.0pt,mark size=2.0pt,mark=x,mark options={solid},forget plot]
 plot [error bars/.cd, y dir = both, y explicit]
 table[row sep=crcr, y error plus index=2, y error minus index=3]{
4	1594.28603061188	79.7478910292091	79.7478910292091	\\
5	6209.86966467515	344.216062837806	344.216062837806	\\
6	16597.2495439596	830.286988320702	830.286988320702	\\
7	37731.9656206143	1697.23915023728	1697.23915023728	\\
8	93332.2624036208	6232.78328301018	6232.78328301018	\\
};
\addplot [color=mycolor4,solid,line width=1.0pt,mark size=2.0pt,mark=x,mark options={solid},forget plot]
 plot [error bars/.cd, y dir = both, y explicit]
 table[row sep=crcr, y error plus index=2, y error minus index=3]{
4	2001.40772299066	114.850718197942	114.850718197942	\\
5	6881.77598166867	342.104736753729	342.104736753729	\\
6	13740.8097855825	647.371712241507	647.371712241507	\\
7	36471.2011203921	1652.69985250508	1652.69985250508	\\
8	73526.6085496023	3887.45915894817	3887.45915894817	\\
};
\addplot [color=mycolor5,solid,line width=1.0pt,mark size=2.0pt,mark=x,mark options={solid},forget plot]
 plot [error bars/.cd, y dir = both, y explicit]
 table[row sep=crcr, y error plus index=2, y error minus index=3]{
4	1913.74822491774	70.1690395034782	70.1690395034782	\\
5	5490.80652654048	204.459537916658	204.459537916658	\\
6	11102.2789044065	361.443144852577	361.443144852577	\\
7	30613.2530069158	1367.07872480957	1367.07872480957	\\
8	63565.3777411037	3148.05248615897	3148.05248615897	\\
};
\addplot [color=mycolor6,solid,line width=1.0pt,mark size=2.0pt,mark=x,mark options={solid},forget plot]
 plot [error bars/.cd, y dir = both, y explicit]
 table[row sep=crcr, y error plus index=2, y error minus index=3]{
4	1732.79055097017	58.1804609307092	58.1804609307092	\\
5	5154.84369128052	214.188638286249	214.188638286249	\\
6	11580.2838685218	527.864897528141	527.864897528141	\\
7	25074.7918727676	940.931869098782	940.931869098782	\\
8	51939.8221582251	2450.25493163793	2450.25493163793	\\
};
\addplot [color=mycolor7,solid,line width=1.0pt,mark size=2.0pt,mark=x,mark options={solid},forget plot]
 plot [error bars/.cd, y dir = both, y explicit]
 table[row sep=crcr, y error plus index=2, y error minus index=3]{
4	1535.12796659363	54.0543997932054	54.0543997932054	\\
5	4628.43385124226	159.007470807302	159.007470807302	\\
6	10165.1697818783	398.698927267495	398.698927267495	\\
7	23356.1951447453	937.428661671733	937.428661671733	\\
8	47726.2163760263	1721.46258995139	1721.46258995139	\\
};
\addplot [color=mycolor8,solid,line width=1.0pt,mark size=2.0pt,mark=x,mark options={solid},forget plot]
 plot [error bars/.cd, y dir = both, y explicit]
 table[row sep=crcr, y error plus index=2, y error minus index=3]{
4	1592.76804017532	44.5173125411898	44.5173125411898	\\
5	4263.46578260356	128.38091300192	128.38091300192	\\
6	9528.00171230492	293.918566738696	293.918566738696	\\
7	19633.9476892859	665.460293698296	665.460293698296	\\
8	41520.4024421473	1454.73084705444	1454.73084705444	\\
};
\end{axis}
\end{tikzpicture}%
}
\caption{{\bf Median time to solution per size.} Shown is the median time to solution for the D-Wave processor (left), HFS (middle), and SAS (right).  The top and bottom rows show data for range-2 and range-3 instances respectively.  Following Hen et al.\ \cite{henfl}, we divide HFS times by $L=\sqrt{N/8}$ to simulate hypothetical parallelization.  SAS data incorporates full $n$-core hypothetical parallelization.   Error bars represent one standard deviation from bootstrap samples; most are smaller than the data markers.}\label{fig:tts}
\end{figure}

\begin{figure}
\begin{center}
\scalebox{.70}{%
\setlength{\figurewidth}{3cm}%
\setlength{\figureheight}{6cm}%
%
%
%
\definecolor{mycolor1}{rgb}{0.00000,0.23333,0.70000}%
\definecolor{mycolor2}{rgb}{0.00000,0.46667,0.70000}%
\definecolor{mycolor3}{rgb}{0.00000,0.70000,0.70000}%
\definecolor{mycolor4}{rgb}{0.23333,0.70000,0.46667}%
\definecolor{mycolor5}{rgb}{0.46667,0.70000,0.23333}%
\definecolor{mycolor6}{rgb}{0.70000,0.70000,0.00000}%
\definecolor{mycolor7}{rgb}{0.70000,0.46667,0.00000}%
\definecolor{mycolor8}{rgb}{0.70000,0.23333,0.00000}%
\begin{tikzpicture}

\begin{axis}[%
width=\figurewidth,
height=\figureheight,
scale only axis,
xmin=3.8,
xmax=8.2,
xlabel={Problem size},
xmajorgrids,
ymode=log,
ymin=0.2,
ymax=200,
yminorticks=true,
ylabel={D-Wave time / HFS time},
ymajorgrids,
title={D-Wave vs. HFS, range 2},
legend style={font=\footnotesize},xtick={4,5,6,7,8},xticklabels={$C_4$,$C_5$,$C_6$,$C_7$,$C_8$}
]
\addplot [color=black!30!blue,solid,line width=1.0pt,mark size=2.0pt,mark=x,mark options={solid},forget plot]
 plot [error bars/.cd, y dir = both, y explicit]
 table[row sep=crcr, y error plus index=2, y error minus index=3]{
4	0.743783336369347	0.0109249295221104	0.0109249295221104	\\
5	1.30972098121643	0.0604946302191824	0.0604946302191824	\\
6	1.65671192106509	0.0935278499028476	0.0935278499028476	\\
7	2.48793710294813	0.180468985938541	0.180468985938541	\\
8	2.20424164436115	0.177378546283528	0.177378546283528	\\
};
\addplot [color=mycolor1,solid,line width=1.0pt,mark size=2.0pt,mark=x,mark options={solid},forget plot]
 plot [error bars/.cd, y dir = both, y explicit]
 table[row sep=crcr, y error plus index=2, y error minus index=3]{
4	1.22751175326009	0.0851172926331061	0.0851172926331061	\\
5	1.56796765169941	0.0759276186955566	0.0759276186955566	\\
6	2.4273035745633	0.13780638445229	0.13780638445229	\\
7	2.82756358959034	0.145191497881486	0.145191497881486	\\
8	4.60218968788039	0.327210809549964	0.327210809549964	\\
};
\addplot [color=mycolor2,solid,line width=1.0pt,mark size=2.0pt,mark=x,mark options={solid},forget plot]
 plot [error bars/.cd, y dir = both, y explicit]
 table[row sep=crcr, y error plus index=2, y error minus index=3]{
4	1.65697917373258	0.0872908072192091	0.0872908072192091	\\
5	2.71904298237916	0.145727579894455	0.145727579894455	\\
6	3.11602902499386	0.263623161379284	0.263623161379284	\\
7	4.8576336556505	0.297886078516289	0.297886078516289	\\
8	6.75907118321325	0.454532691531647	0.454532691531647	\\
};
\addplot [color=mycolor3,solid,line width=2.0pt,mark size=2.0pt,mark=x,mark options={solid},forget plot]
 plot [error bars/.cd, y dir = both, y explicit]
 table[row sep=crcr, y error plus index=2, y error minus index=3]{
4	1.51977991033761	0.0921062760712221	0.0921062760712221	\\
5	2.61252696465164	0.218178720712512	0.218178720712512	\\
6	4.00130939099564	0.23888865094285	0.238888650942851	\\
7	6.89714156804008	0.404475109351204	0.404475109351204	\\
8	9.19035217250158	0.607140920743417	0.607140920743417	\\
};
\addplot [color=mycolor4,solid,line width=1.0pt,mark size=2.0pt,mark=x,mark options={solid},forget plot]
 plot [error bars/.cd, y dir = both, y explicit]
 table[row sep=crcr, y error plus index=2, y error minus index=3]{
4	1.7996939370086	0.0623177358711575	0.0623177358711575	\\
5	3.35124555065048	0.15250898720239	0.15250898720239	\\
6	5.63920468873154	0.431546968829528	0.431546968829528	\\
7	8.30954454930403	0.534874731787577	0.534874731787577	\\
8	13.1137499676745	0.706769560386173	0.706769560386173	\\
};
\addplot [color=mycolor5,solid,line width=1.0pt,mark size=2.0pt,mark=x,mark options={solid},forget plot]
 plot [error bars/.cd, y dir = both, y explicit]
 table[row sep=crcr, y error plus index=2, y error minus index=3]{
4	1.80559852090325	0.108977175177835	0.108977175177835	\\
5	3.96556844069518	0.148810979793313	0.148810979793313	\\
6	6.31337386143653	0.294897833470181	0.294897833470181	\\
7	9.70640363712361	0.532769304974968	0.532769304974968	\\
8	17.5933205778032	0.858177449226176	0.858177449226176	\\
};
\addplot [color=mycolor6,solid,line width=1.0pt,mark size=2.0pt,mark=x,mark options={solid},forget plot]
 plot [error bars/.cd, y dir = both, y explicit]
 table[row sep=crcr, y error plus index=2, y error minus index=3]{
4	1.23851047517747	0.054073205536177	0.054073205536177	\\
5	3.30786937446634	0.128625885411943	0.128625885411943	\\
6	7.00404371144306	0.238302047179555	0.238302047179555	\\
7	9.78116624805965	0.550154569728647	0.550154569728647	\\
8	19.7499710901778	0.913571906655804	0.913571906655804	\\
};
\addplot [color=mycolor7,solid,line width=1.0pt,mark size=2.0pt,mark=x,mark options={solid},forget plot]
 plot [error bars/.cd, y dir = both, y explicit]
 table[row sep=crcr, y error plus index=2, y error minus index=3]{
4	1.31907522302816	0.0653453674776787	0.0653453674776787	\\
5	2.8562168325825	0.0939491650015656	0.0939491650015656	\\
6	6.19042562448857	0.189538225570393	0.189538225570393	\\
7	10.8975588940899	0.343592208436659	0.343592208436659	\\
8	19.2186931033506	0.618294433737471	0.618294433737471	\\
};
\addplot [color=mycolor8,solid,line width=1.0pt,mark size=2.0pt,mark=x,mark options={solid},forget plot]
 plot [error bars/.cd, y dir = both, y explicit]
 table[row sep=crcr, y error plus index=2, y error minus index=3]{
4	1.24883809264146	0.0083724886839629	0.0083724886839629	\\
5	3.1316220658879	0.120893422112232	0.120893422112232	\\
6	5.67307776408264	0.201819781379903	0.201819781379903	\\
7	10.1218968061212	0.286740033090732	0.286740033090732	\\
8	19.0237299608636	0.565517821600253	0.565517821600253	\\
};
\end{axis}
\end{tikzpicture}%
\hspace{.5cm}
%
%
%
\definecolor{mycolor1}{rgb}{0.00000,0.23333,0.70000}%
\definecolor{mycolor2}{rgb}{0.00000,0.46667,0.70000}%
\definecolor{mycolor3}{rgb}{0.00000,0.70000,0.70000}%
\definecolor{mycolor4}{rgb}{0.23333,0.70000,0.46667}%
\definecolor{mycolor5}{rgb}{0.46667,0.70000,0.23333}%
\definecolor{mycolor6}{rgb}{0.70000,0.70000,0.00000}%
\definecolor{mycolor7}{rgb}{0.70000,0.46667,0.00000}%
\definecolor{mycolor8}{rgb}{0.70000,0.23333,0.00000}%
\begin{tikzpicture}

\begin{axis}[%
width=\figurewidth,
height=\figureheight,
scale only axis,
xmin=3.8,
xmax=8.2,
xlabel={Problem size},
xmajorgrids,
ymode=log,
ymin=0.2,
ymax=200,
yminorticks=true,
ymajorgrids,
title={D-Wave vs. HFS, range 3},
legend style={at={(1.03,0.5)},anchor=west,fill=none,draw=none,legend cell align=left},
legend style={font=\footnotesize},xtick={4,5,6,7,8},xticklabels={$C_4$,$C_5$,$C_6$,$C_7$,$C_8$}
]
\addplot [color=black!30!blue,solid,line width=1.0pt,mark size=2.0pt,mark=x,mark options={solid}]
 plot [error bars/.cd, y dir = both, y explicit]
 table[row sep=crcr, y error plus index=2, y error minus index=3]{
4	0.571948590213589	0.0504344135889681	0.0504344135889681	\\
5	0.840478160369835	0.0628136229567031	0.0628136229567031	\\
6	1.16306347588439	0.0897769433696667	0.0897769433696667	\\
7	1.21176143890176	0.0924357238204494	0.0924357238204494	\\
8	1.20846179159664	0.0716602979274379	0.0716602979274379	\\
};
\addlegendentry{$\alpha = 0.10$};

\addplot [color=mycolor1,solid,line width=1.0pt,mark size=2.0pt,mark=x,mark options={solid}]
 plot [error bars/.cd, y dir = both, y explicit]
 table[row sep=crcr, y error plus index=2, y error minus index=3]{
4	0.844591081298399	0.0705439497472781	0.0705439497472781	\\
5	1.02252798978141	0.0558448424313081	0.0558448424313081	\\
6	1.20549031841518	0.0897168538518491	0.0897168538518491	\\
7	1.6064526749514	0.0936846478633906	0.0936846478633906	\\
8	1.39974366090659	0.107041283554927	0.107041283554927	\\
};
\addlegendentry{$\alpha = 0.15$};

\addplot [color=mycolor2,solid,line width=1.0pt,mark size=2.0pt,mark=x,mark options={solid}]
 plot [error bars/.cd, y dir = both, y explicit]
 table[row sep=crcr, y error plus index=2, y error minus index=3]{
4	1.09906415097629	0.0723512360378342	0.0723512360378342	\\
5	1.31258717886305	0.0832660029268479	0.0832660029268479	\\
6	1.5288509597017	0.0924473349809805	0.0924473349809805	\\
7	2.47914509669516	0.142403343319547	0.142403343319547	\\
8	2.5860063142552	0.203004759681491	0.203004759681491	\\
};
\addlegendentry{$\alpha = 0.20$};

\addplot [color=mycolor3,solid,line width=2.0pt,mark size=2.0pt,mark=x,mark options={solid}]
 plot [error bars/.cd, y dir = both, y explicit]
 table[row sep=crcr, y error plus index=2, y error minus index=3]{
4	1.20941180333332	0.07231077055039	0.07231077055039	\\
5	1.39594255348531	0.0676426814931712	0.0676426814931712	\\
6	2.28059331193565	0.131335271354376	0.131335271354376	\\
7	2.81681642029031	0.171822047370205	0.171822047370205	\\
8	3.82708307424392	0.252535951312397	0.252535951312397	\\
};
\addlegendentry{$\alpha = 0.25$};

\addplot [color=mycolor4,solid,line width=1.0pt,mark size=2.0pt,mark=x,mark options={solid}]
 plot [error bars/.cd, y dir = both, y explicit]
 table[row sep=crcr, y error plus index=2, y error minus index=3]{
4	1.24945858019459	0.0569842047279046	0.0569842047279046	\\
5	2.02582718106465	0.153107250486877	0.153107250486877	\\
6	3.15540582085941	0.225141143761876	0.225141143761876	\\
7	4.03693152370519	0.273002345576664	0.273002345576665	\\
8	5.59993441425673	0.384287760262381	0.384287760262381	\\
};
\addlegendentry{$\alpha = 0.30$};

\addplot [color=mycolor5,solid,line width=1.0pt,mark size=2.0pt,mark=x,mark options={solid}]
 plot [error bars/.cd, y dir = both, y explicit]
 table[row sep=crcr, y error plus index=2, y error minus index=3]{
4	1.49702564497203	0.0479566301199845	0.0479566301199845	\\
5	2.77554434150828	0.167631288980942	0.167631288980942	\\
6	3.99255173677022	0.222203153897327	0.222203153897328	\\
7	5.66670208267986	0.358008076396779	0.358008076396779	\\
8	8.00039042136417	0.490813382891988	0.490813382891988	\\
};
\addlegendentry{$\alpha = 0.35$};

\addplot [color=mycolor6,solid,line width=1.0pt,mark size=2.0pt,mark=x,mark options={solid}]
 plot [error bars/.cd, y dir = both, y explicit]
 table[row sep=crcr, y error plus index=2, y error minus index=3]{
4	1.49817701368505	0.058925613834184	0.058925613834184	\\
5	3.00786117367006	0.143775183128017	0.143775183128017	\\
6	4.46840913344986	0.252562491307357	0.252562491307357	\\
7	6.83646098507278	0.406750724440527	0.406750724440527	\\
8	10.5980089081586	0.843469941876421	0.843469941876421	\\
};
\addlegendentry{$\alpha = 0.40$};

\addplot [color=mycolor7,solid,line width=1.0pt,mark size=2.0pt,mark=x,mark options={solid}]
 plot [error bars/.cd, y dir = both, y explicit]
 table[row sep=crcr, y error plus index=2, y error minus index=3]{
4	1.30166906392547	0.118722564891001	0.118722564891001	\\
5	3.1178626621207	0.132095513848038	0.132095513848038	\\
6	5.46970279534348	0.203175140169084	0.203175140169084	\\
7	7.79499239277287	0.460530068868274	0.460530068868274	\\
8	12.9636616775155	0.969505756589323	0.969505756589323	\\
};
\addlegendentry{$\alpha = 0.45$};

\addplot [color=mycolor8,solid,line width=1.0pt,mark size=2.0pt,mark=x,mark options={solid}]
 plot [error bars/.cd, y dir = both, y explicit]
 table[row sep=crcr, y error plus index=2, y error minus index=3]{
4	1.21220309144749	0.0722093211284118	0.0722093211284118	\\
5	3.18347224994617	0.0897745790091733	0.0897745790091733	\\
6	5.47480655745816	0.235235882192397	0.235235882192397	\\
7	8.55991036719221	0.384788926708518	0.384788926708518	\\
8	13.6189756625424	0.616267623227071	0.616267623227071	\\
};
\addlegendentry{$\alpha = 0.50$};

\end{axis}
\end{tikzpicture}%
}\vspace{.2cm}
\scalebox{.70}{%
\setlength{\figurewidth}{3cm}%
\setlength{\figureheight}{6cm}%
%
%
%
\definecolor{mycolor1}{rgb}{0.00000,0.23333,0.70000}%
\definecolor{mycolor2}{rgb}{0.00000,0.46667,0.70000}%
\definecolor{mycolor3}{rgb}{0.00000,0.70000,0.70000}%
\definecolor{mycolor4}{rgb}{0.23333,0.70000,0.46667}%
\definecolor{mycolor5}{rgb}{0.46667,0.70000,0.23333}%
\definecolor{mycolor6}{rgb}{0.70000,0.70000,0.00000}%
\definecolor{mycolor7}{rgb}{0.70000,0.46667,0.00000}%
\definecolor{mycolor8}{rgb}{0.70000,0.23333,0.00000}%
\begin{tikzpicture}

\begin{axis}[%
width=\figurewidth,
height=\figureheight,
scale only axis,
xmin=3.8,
xmax=8.2,
xlabel={Problem size},
xmajorgrids,
ymode=log,
ymin=2,
ymax=2000,
yminorticks=true,
ylabel={D-Wave time / SAS time},
ymajorgrids,
title={D-Wave vs. SAS, range 2},
legend style={font=\footnotesize},xtick={4,5,6,7,8},xticklabels={$C_4$,$C_5$,$C_6$,$C_7$,$C_8$}
]
\addplot [color=black!30!blue,solid,line width=1.0pt,mark size=2.0pt,mark=x,mark options={solid},forget plot]
 plot [error bars/.cd, y dir = both, y explicit]
 table[row sep=crcr, y error plus index=2, y error minus index=3]{
4	4.96420590323267	0.21307788382997	0.21307788382997	\\
5	8.81882230500344	0.971313008887394	0.971313008887393	\\
6	15.1568642457419	1.5350367461072	1.5350367461072	\\
7	19.9097711208916	1.96984291187914	1.96984291187914	\\
8	29.9785801442278	3.51269052116842	3.51269052116842	\\
};
\addplot [color=mycolor1,solid,line width=1.0pt,mark size=2.0pt,mark=x,mark options={solid},forget plot]
 plot [error bars/.cd, y dir = both, y explicit]
 table[row sep=crcr, y error plus index=2, y error minus index=3]{
4	13.4066973623929	1.5733236966478	1.5733236966478	\\
5	20.6640877719061	1.09043285597468	1.09043285597468	\\
6	40.7179634796554	3.81841711074665	3.81841711074665	\\
7	65.7591101003381	3.88402926097474	3.88402926097474	\\
8	86.5339830204569	7.93238534674813	7.93238534674813	\\
};
\addplot [color=mycolor2,solid,line width=1.0pt,mark size=2.0pt,mark=x,mark options={solid},forget plot]
 plot [error bars/.cd, y dir = both, y explicit]
 table[row sep=crcr, y error plus index=2, y error minus index=3]{
4	27.8893299638107	1.75397670585113	1.75397670585113	\\
5	62.907200415094	4.36729897394946	4.36729897394946	\\
6	105.239041797764	12.2435120009113	12.2435120009113	\\
7	160.048182191125	16.9030399136163	16.9030399136163	\\
8	192.652759955138	17.8342851049513	17.8342851049513	\\
};
\addplot [color=mycolor3,solid,line width=2.0pt,mark size=2.0pt,mark=x,mark options={solid},forget plot]
 plot [error bars/.cd, y dir = both, y explicit]
 table[row sep=crcr, y error plus index=2, y error minus index=3]{
4	43.6384378845646	2.78915604745058	2.78915604745058	\\
5	77.9920560506847	7.26656450269593	7.26656450269593	\\
6	126.878380266569	8.19788566573321	8.19788566573321	\\
7	230.224419354011	15.8062761737714	15.8062761737714	\\
8	306.283477826561	29.8890631468679	29.8890631468679	\\
};
\addplot [color=mycolor4,solid,line width=1.0pt,mark size=2.0pt,mark=x,mark options={solid},forget plot]
 plot [error bars/.cd, y dir = both, y explicit]
 table[row sep=crcr, y error plus index=2, y error minus index=3]{
4	59.6321644604549	2.69981822927976	2.69981822927976	\\
5	132.521775306387	8.73898261250056	8.73898261250056	\\
6	210.848301290259	16.3784737455937	16.3784737455937	\\
7	315.844933786387	24.3533776230744	24.3533776230744	\\
8	419.824184695802	28.3144666510019	28.3144666510019	\\
};
\addplot [color=mycolor5,solid,line width=1.0pt,mark size=2.0pt,mark=x,mark options={solid},forget plot]
 plot [error bars/.cd, y dir = both, y explicit]
 table[row sep=crcr, y error plus index=2, y error minus index=3]{
4	66.2040719221387	2.72541184657395	2.72541184657395	\\
5	157.741953286639	7.01151716606091	7.01151716606091	\\
6	239.968195184661	15.3808668827049	15.3808668827049	\\
7	356.22123803142	21.3035677094429	21.3035677094429	\\
8	565.549133499842	39.0303903477817	39.0303903477817	\\
};
\addplot [color=mycolor6,solid,line width=1.0pt,mark size=2.0pt,mark=x,mark options={solid},forget plot]
 plot [error bars/.cd, y dir = both, y explicit]
 table[row sep=crcr, y error plus index=2, y error minus index=3]{
4	81.0117840479437	2.87062911212672	2.87062911212672	\\
5	150.950038716879	3.34027454672585	3.34027454672585	\\
6	287.704221462442	9.65300930051058	9.65300930051058	\\
7	398.05539615486	25.6453388780793	25.6453388780793	\\
8	720.384182669538	42.2282869762184	42.2282869762184	\\
};
\addplot [color=mycolor7,solid,line width=1.0pt,mark size=2.0pt,mark=x,mark options={solid},forget plot]
 plot [error bars/.cd, y dir = both, y explicit]
 table[row sep=crcr, y error plus index=2, y error minus index=3]{
4	83.9874200358164	2.09807735121707	2.09807735121707	\\
5	158.568138321883	5.0402248151747	5.0402248151747	\\
6	332.024991873052	11.3933211909222	11.3933211909222	\\
7	563.997435360169	21.3627798331305	21.3627798331305	\\
8	840.379629951882	31.4893305096392	31.4893305096392	\\
};
\addplot [color=mycolor8,solid,line width=1.0pt,mark size=2.0pt,mark=x,mark options={solid},forget plot]
 plot [error bars/.cd, y dir = both, y explicit]
 table[row sep=crcr, y error plus index=2, y error minus index=3]{
4	88.5965011264527	1.85597292959065	1.85597292959065	\\
5	173.964313900363	4.74536617888606	4.74536617888606	\\
6	306.89695911972	6.37057897367828	6.37057897367828	\\
7	556.83437448789	12.9548703851503	12.9548703851503	\\
8	998.251193831365	23.971096739132	23.971096739132	\\
};
\end{axis}
\end{tikzpicture}%
\hspace{.5cm}
%
%
%
\definecolor{mycolor1}{rgb}{0.00000,0.23333,0.70000}%
\definecolor{mycolor2}{rgb}{0.00000,0.46667,0.70000}%
\definecolor{mycolor3}{rgb}{0.00000,0.70000,0.70000}%
\definecolor{mycolor4}{rgb}{0.23333,0.70000,0.46667}%
\definecolor{mycolor5}{rgb}{0.46667,0.70000,0.23333}%
\definecolor{mycolor6}{rgb}{0.70000,0.70000,0.00000}%
\definecolor{mycolor7}{rgb}{0.70000,0.46667,0.00000}%
\definecolor{mycolor8}{rgb}{0.70000,0.23333,0.00000}%
\begin{tikzpicture}

\begin{axis}[%
width=\figurewidth,
height=\figureheight,
scale only axis,
xmin=3.8,
xmax=8.2,
xlabel={Problem size},
xmajorgrids,
ymode=log,
ymin=2,
ymax=2000,
yminorticks=true,
ymajorgrids,
title={D-Wave vs. SAS, range 3},
legend style={at={(1.03,0.5)},anchor=west,fill=none,draw=none,legend cell align=left},
legend style={font=\footnotesize},xtick={4,5,6,7,8},xticklabels={$C_4$,$C_5$,$C_6$,$C_7$,$C_8$}
]
\addplot [color=black!30!blue,solid,line width=1.0pt,mark size=2.0pt,mark=x,mark options={solid}]
 plot [error bars/.cd, y dir = both, y explicit]
 table[row sep=crcr, y error plus index=2, y error minus index=3]{
4	4.55032888579965	0.458767558098448	0.458767558098448	\\
5	6.99173415339468	1.1043702644747	1.1043702644747	\\
6	12.3715622991194	1.47991252352397	1.47991252352397	\\
7	13.5927884958786	1.90328046511046	1.90328046511046	\\
8	15.9349854582491	1.08977698487437	1.08977698487437	\\
};
\addlegendentry{$\alpha = 0.10$};

\addplot [color=mycolor1,solid,line width=1.0pt,mark size=2.0pt,mark=x,mark options={solid}]
 plot [error bars/.cd, y dir = both, y explicit]
 table[row sep=crcr, y error plus index=2, y error minus index=3]{
4	13.4437783946176	1.5239742211568	1.5239742211568	\\
5	16.1843136457342	0.995689655630944	0.995689655630946	\\
6	22.5269418970612	2.08066558945531	2.08066558945531	\\
7	26.0694490165088	2.16181371250554	2.16181371250554	\\
8	29.706681810574	3.809885599051	3.809885599051	\\
};
\addlegendentry{$\alpha = 0.15$};

\addplot [color=mycolor2,solid,line width=1.0pt,mark size=2.0pt,mark=x,mark options={solid}]
 plot [error bars/.cd, y dir = both, y explicit]
 table[row sep=crcr, y error plus index=2, y error minus index=3]{
4	19.6166599316596	1.34549550250998	1.34549550250998	\\
5	28.7723254776881	2.55482078917252	2.55482078917252	\\
6	37.4800219401354	3.20395853151202	3.20395853151202	\\
7	61.1910255980744	5.27449107950547	5.27449107950547	\\
8	63.047310896303	6.48325267540692	6.48325267540692	\\
};
\addlegendentry{$\alpha = 0.20$};

\addplot [color=mycolor3,solid,line width=2.0pt,mark size=2.0pt,mark=x,mark options={solid}]
 plot [error bars/.cd, y dir = both, y explicit]
 table[row sep=crcr, y error plus index=2, y error minus index=3]{
4	27.5442077330557	1.71599049935408	1.71599049935408	\\
5	42.2824857481677	2.87254187026607	2.87254187026607	\\
6	63.65974386437	4.55901593403863	4.55901593403863	\\
7	90.0683857320626	6.1866713103112	6.1866713103112	\\
8	113.798431297122	10.8505423694777	10.8505423694777	\\
};
\addlegendentry{$\alpha = 0.25$};

\addplot [color=mycolor4,solid,line width=1.0pt,mark size=2.0pt,mark=x,mark options={solid}]
 plot [error bars/.cd, y dir = both, y explicit]
 table[row sep=crcr, y error plus index=2, y error minus index=3]{
4	41.2020935685966	2.92957545081696	2.92957545081696	\\
5	63.6874547016848	5.46459886377629	5.46459886377629	\\
6	94.7763418946878	7.92700592226674	7.92700592226674	\\
7	129.573711565731	10.0025166131848	10.0025166131848	\\
8	158.708264023122	13.114055110656	13.114055110656	\\
};
\addlegendentry{$\alpha = 0.30$};

\addplot [color=mycolor5,solid,line width=1.0pt,mark size=2.0pt,mark=x,mark options={solid}]
 plot [error bars/.cd, y dir = both, y explicit]
 table[row sep=crcr, y error plus index=2, y error minus index=3]{
4	47.7482559065827	2.09260398402304	2.09260398402304	\\
5	82.4476528315695	5.01248840260301	5.01248840260301	\\
6	113.559216200318	7.02155827082323	7.02155827082323	\\
7	184.888197455228	13.7759778413484	13.7759778413484	\\
8	232.83585492153	17.9367336023551	17.9367336023551	\\
};
\addlegendentry{$\alpha = 0.35$};

\addplot [color=mycolor6,solid,line width=1.0pt,mark size=2.0pt,mark=x,mark options={solid}]
 plot [error bars/.cd, y dir = both, y explicit]
 table[row sep=crcr, y error plus index=2, y error minus index=3]{
4	50.1944432633218	1.83675107178763	1.83675107178763	\\
5	104.608021642645	5.90235131343864	5.90235131343864	\\
6	140.541576498547	9.39456572891606	9.39456572891606	\\
7	232.337989635356	14.205527720219	14.205527720219	\\
8	306.734893034949	28.3947704328821	28.3947704328821	\\
};
\addlegendentry{$\alpha = 0.40$};

\addplot [color=mycolor7,solid,line width=1.0pt,mark size=2.0pt,mark=x,mark options={solid}]
 plot [error bars/.cd, y dir = both, y explicit]
 table[row sep=crcr, y error plus index=2, y error minus index=3]{
4	50.4553191184978	1.79239768893004	1.79239768893004	\\
5	109.813832565755	4.51426792207222	4.51426792207222	\\
6	192.287232328356	8.71577182123374	8.71577182123374	\\
7	272.633582878669	18.137528988006	18.137528988006	\\
8	414.429518873099	35.4957067919882	35.4957067919882	\\
};
\addlegendentry{$\alpha = 0.45$};

\addplot [color=mycolor8,solid,line width=1.0pt,mark size=2.0pt,mark=x,mark options={solid}]
 plot [error bars/.cd, y dir = both, y explicit]
 table[row sep=crcr, y error plus index=2, y error minus index=3]{
4	56.4400594826818	1.55976827519623	1.55976827519623	\\
5	118.378201141433	3.75415533328162	3.75415533328162	\\
6	199.597590171038	8.77019947060089	8.77019947060089	\\
7	325.730373626634	17.1630250964408	17.1630250964408	\\
8	479.635621863916	25.5078547553632	25.5078547553632	\\
};
\addlegendentry{$\alpha = 0.50$};

\end{axis}
\end{tikzpicture}%
}
\caption{{\bf Median ratio of running time by size.} Shown is the median ratio of time to solution for the D-Wave processor compared with each classical solver.  Positive slope represents an increasing advantage for the D-Wave processor as problems get larger.  The hardest overall regime roughly corresponds to $\alpha \approx 0.25$.  Error bars represent one standard deviation from bootstrap samples.}\label{fig:speedup}
\end{center}
\end{figure}
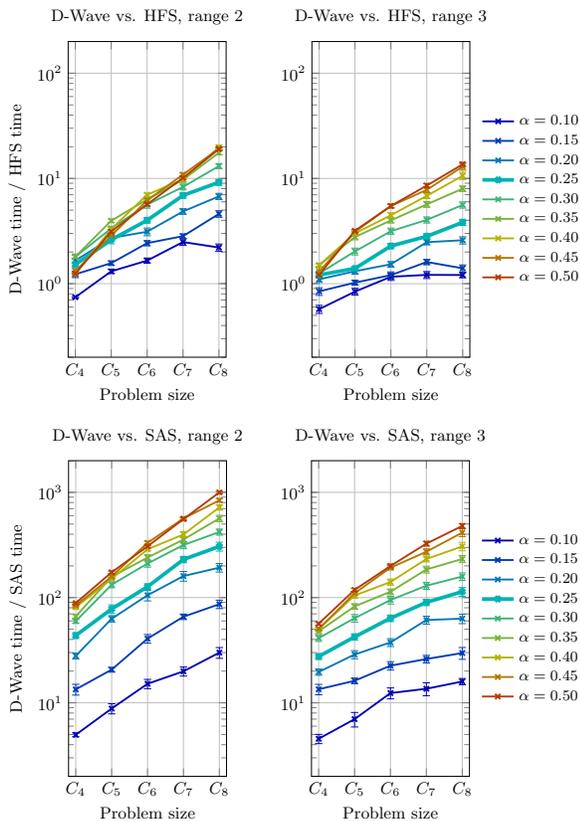


\subsection{Performance scaling results}

In Fig.\ \ref{fig:tts} we show the scaling of the median time to solution for the three solvers as the problem size $L$ increases.  As in previous work \cite{henfl,Roennow2014}, we are particularly interested in how the ratio between two solvers' time to solution scales with respect to problem size.  This is shown in Fig.\ \ref{fig:speedup}.  A positive slope in Fig.\ \ref{fig:speedup} indicates a performance scaling advantage for the D-Wave processor, and the possibility of {\em limited quantum speedup} as defined by R{\o}nnow et al.\ \cite{Roennow2014} in the case of SAS, and the possibility of {\em potential quantum speedup} in the case of HFS\footnote{The distinction arises because HFS is a combinatorial algorithm rather than one based on a physical model \cite{henfl}.}.  In the Supplemental Material \cite{supp} we arrange the data for range-2, range-3, and range-unlimited instances by $\alpha$.

D-Wave Vesuvius processors allow a minimum anneal time of $20\si{\micro\second}$; previous work has shown that $C_8$-scale problems with optimal anneal time greater than this are elusive \cite{henfl, Roennow2014, Venturelli2014}.  Proving limited quantum speedup in this framework would require data from the D-Wave processor using shorter anneals to certify that we are not artificially slowing the processor on easier instances.  In particular for the smaller and easier problems, the minimum anneal time may mask the true performance scaling of the quantum annealing platform \cite{henfl, Amin2015}.  This may explain, to some extent, the outstanding performance of the D-Wave processor on high-$\alpha$ instances.  

In the Supplemental Material \cite{supp} we analyze the effect that range limitation has on the difficulty of the problems.  Here we simply note that for the hardest range of $\alpha$, limiting the range of instances to 3 does not seem to make the problems significantly easier.  This can be seen where range-2, range-3, and range-unlimited instances are compared for the available solvers.

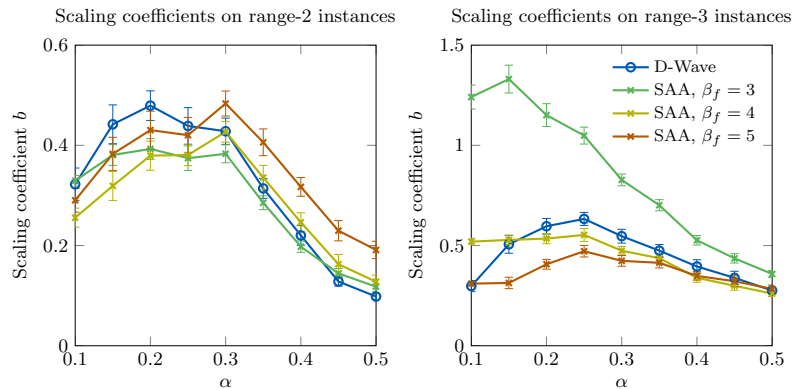
\begin{figure*}
\scalebox{.8}{%
\setlength{\figurewidth}{5cm}%
\setlength{\figureheight}{5cm}%
%
%
%
\definecolor{mycolor1}{rgb}{0.00000,0.35000,0.70000}%
\definecolor{mycolor2}{rgb}{0.35000,0.70000,0.35000}%
\definecolor{mycolor3}{rgb}{0.70000,0.70000,0.00000}%
\begin{tikzpicture}

\begin{axis}[%
width=\figurewidth,
height=\figureheight,
scale only axis,
xmin=0.1,
xmax=0.5,
xlabel={$\alpha$},
ymin=0,
ymax=0.6,
ylabel={Scaling coefficient $b$},
title={Scaling coefficients on range-2 instances},
legend style={font=\footnotesize},
]
\addplot [color=mycolor1,solid,line width=1.0pt,mark size=2.0pt,mark=o,mark options={solid},forget plot]
 plot [error bars/.cd, y dir = both, y explicit]
 table[row sep=crcr, y error plus index=2, y error minus index=3]{
0.1	0.322128254748371	0.0325497275906554	0.0325497275906554	\\
0.15	0.441912441264473	0.0387596251956538	0.0387596251956538	\\
0.2	0.479248019550589	0.0295594159324041	0.0295594159324041	\\
0.25	0.438677626182446	0.0366242231370257	0.0366242231370257	\\
0.3	0.428092024305453	0.0261844581020746	0.0261844581020746	\\
0.35	0.313997604441917	0.0199180193751765	0.0199180193751765	\\
0.4	0.220194205317284	0.0196163250702769	0.0196163250702769	\\
0.45	0.128231504576093	0.00918654547954503	0.00918654547954503	\\
0.5	0.0981954984510208	0.0054063662353077	0.0054063662353077	\\
};
\addplot [color=mycolor2,solid,line width=1.0pt,mark size=2.0pt,mark=x,mark options={solid},forget plot]
 plot [error bars/.cd, y dir = both, y explicit]
 table[row sep=crcr, y error plus index=2, y error minus index=3]{
0.1	0.329349658540206	0.0107075065409394	0.0107075065409394	\\
0.15	0.380713096978496	0.0207498824568836	0.0207498824568836	\\
0.2	0.392789910797293	0.0207282861609442	0.0207282861609442	\\
0.25	0.37427110578217	0.0246150894483708	0.0246150894483708	\\
0.3	0.383100445011558	0.0179431709408941	0.0179431709408941	\\
0.35	0.28536869405411	0.0137328549407211	0.0137328549407211	\\
0.4	0.197806556022198	0.0112826848528881	0.0112826848528881	\\
0.45	0.144326032418011	0.0099043631129066	0.0099043631129066	\\
0.5	0.117852418706913	0.00991099126635434	0.00991099126635434	\\
};
\addplot [color=mycolor3,solid,line width=1.0pt,mark size=2.0pt,mark=x,mark options={solid},forget plot]
 plot [error bars/.cd, y dir = both, y explicit]
 table[row sep=crcr, y error plus index=2, y error minus index=3]{
0.1	0.255540905731153	0.0190741935763666	0.0190741935763667	\\
0.15	0.319108325942217	0.0290693561924462	0.0290693561924462	\\
0.2	0.379460429459289	0.0290695104609343	0.0290695104609343	\\
0.25	0.380677163756149	0.0212855255899215	0.0212855255899215	\\
0.3	0.426958704671497	0.0207787709200349	0.0207787709200349	\\
0.35	0.335842676471966	0.0242000311750665	0.0242000311750665	\\
0.4	0.246342392915503	0.018920061020796	0.018920061020796	\\
0.45	0.162745906044727	0.0194066609421513	0.0194066609421513	\\
0.5	0.127816259492426	0.0130292398875331	0.0130292398875331	\\
};
\addplot [color=black!30!orange,solid,line width=1.0pt,mark size=2.0pt,mark=x,mark options={solid},forget plot]
 plot [error bars/.cd, y dir = both, y explicit]
 table[row sep=crcr, y error plus index=2, y error minus index=3]{
0.1	0.29027393739503	0.0236551246328272	0.0236551246328272	\\
0.15	0.382847677647337	0.0332221446685284	0.0332221446685284	\\
0.2	0.430396621215775	0.0376925295882909	0.0376925295882909	\\
0.25	0.420266588887923	0.0354434345437959	0.0354434345437959	\\
0.3	0.483365965427959	0.0249543980294369	0.0249543980294369	\\
0.35	0.406036038409625	0.0266858544298507	0.0266858544298507	\\
0.4	0.317033097724996	0.0187122739516942	0.0187122739516942	\\
0.45	0.2296110598778	0.0202357148242305	0.0202357148242305	\\
0.5	0.191196238437804	0.0173035963306678	0.0173035963306678	\\
};
\end{axis}
\end{tikzpicture}%
%
%
%
\definecolor{mycolor1}{rgb}{0.00000,0.35000,0.70000}%
\definecolor{mycolor2}{rgb}{0.35000,0.70000,0.35000}%
\definecolor{mycolor3}{rgb}{0.70000,0.70000,0.00000}%
\begin{tikzpicture}

\begin{axis}[%
width=\figurewidth,
height=\figureheight,
scale only axis,
xmin=0.1,
xmax=0.5,
xlabel={$\alpha$},
ymin=0,
ymax=1.5,
ylabel={Scaling coefficient $b$},
title={Scaling coefficients on range-3 instances},
legend style={fill=none,draw=none,legend cell align=left},
legend style={font=\footnotesize},
]
\addplot [color=mycolor1,solid,line width=1.0pt,mark size=2.0pt,mark=o,mark options={solid}]
 plot [error bars/.cd, y dir = both, y explicit]
 table[row sep=crcr, y error plus index=2, y error minus index=3]{
0.1	0.297675094277381	0.0278344029995581	0.0278344029995581	\\
0.15	0.50619496698388	0.0444994936449056	0.0444994936449056	\\
0.2	0.596197887415675	0.0388260586109781	0.0388260586109781	\\
0.25	0.632880976475396	0.0325030201746958	0.0325030201746958	\\
0.3	0.546683948403929	0.0343760440636353	0.0343760440636353	\\
0.35	0.474527831842528	0.0304822128141591	0.0304822128141591	\\
0.4	0.395519268182832	0.0337130563925304	0.0337130563925304	\\
0.45	0.338860761900167	0.0327911925328925	0.0327911925328925	\\
0.5	0.275431754139263	0.0172417835999646	0.0172417835999646	\\
};
\addlegendentry{D-Wave};

\addplot [color=mycolor2,solid,line width=1.0pt,mark size=2.0pt,mark=x,mark options={solid}]
 plot [error bars/.cd, y dir = both, y explicit]
 table[row sep=crcr, y error plus index=2, y error minus index=3]{
0.1	1.24098024331606	0.0597950299260352	0.0597950299260352	\\
0.15	1.33058520852893	0.0694078422460516	0.0694078422460516	\\
0.2	1.15076873176639	0.0569557411818089	0.0569557411818089	\\
0.25	1.04854023149273	0.0422500634439276	0.0422500634439276	\\
0.3	0.827448714700228	0.029092679195415	0.029092679195415	\\
0.35	0.701064560675392	0.0278266358832989	0.0278266358832989	\\
0.4	0.526371608325616	0.0235175430185877	0.0235175430185877	\\
0.45	0.4360145146105	0.0232478748261371	0.0232478748261371	\\
0.5	0.358190716523339	0.0168990837832146	0.0168990837832146	\\
};
\addlegendentry{SAA, $\beta_f=3$};

\addplot [color=mycolor3,solid,line width=1.0pt,mark size=2.0pt,mark=x,mark options={solid}]
 plot [error bars/.cd, y dir = both, y explicit]
 table[row sep=crcr, y error plus index=2, y error minus index=3]{
0.1	0.519428701758709	0.0169650113486508	0.0169650113486508	\\
0.15	0.528083619447494	0.0234794921319231	0.0234794921319231	\\
0.2	0.534637676482514	0.0255478585831002	0.0255478585831002	\\
0.25	0.553135000783945	0.0325539732379286	0.0325539732379286	\\
0.3	0.474167889109919	0.0214542625374969	0.0214542625374969	\\
0.35	0.436806067914202	0.0244984571965836	0.0244984571965836	\\
0.4	0.33923726283443	0.0224514944554139	0.0224514944554139	\\
0.45	0.299129388649558	0.0221759793917828	0.0221759793917828	\\
0.5	0.260358266728752	0.014744803820827	0.014744803820827	\\
};
\addlegendentry{SAA, $\beta_f=4$};

\addplot [color=black!30!orange,solid,line width=1.0pt,mark size=2.0pt,mark=x,mark options={solid}]
 plot [error bars/.cd, y dir = both, y explicit]
 table[row sep=crcr, y error plus index=2, y error minus index=3]{
0.1	0.310060642449788	0.0102612263154956	0.0102612263154956	\\
0.15	0.313455961762507	0.0284986280872334	0.0284986280872334	\\
0.2	0.406531496385269	0.0244146339013477	0.0244146339013477	\\
0.25	0.471627056618461	0.0288111832265757	0.0288111832265757	\\
0.3	0.423614443919892	0.0267724047277884	0.0267724047277884	\\
0.35	0.413925365272091	0.0256133062474527	0.0256133062474527	\\
0.4	0.348425308946114	0.0211062620907013	0.0211062620907013	\\
0.45	0.32188448418036	0.0272074097188912	0.0272074097188912	\\
0.5	0.283719464582309	0.0193788147054756	0.0193788147054756	\\
};
\addlegendentry{SAA, $\beta_f=5$};

\end{axis}
\end{tikzpicture}%
}
\caption{{\bf Scaling coefficients for the D-Wave processor and SAA.} Shown are scaling coefficients (exponential slope fits, i.e.\ time to solution $\propto \exp(b(\alpha)L)$) for performance of the D-Wave processor and SAA on range-2 and range-3 instances.  Hen et al.\ find agreement between the D-Wave processor and SAA at $\beta_f=5$.  Here each final inverse temperature $\beta_f\in\{3,4,5\}$ appears deficient in some regime as a model for performance of the D-Wave processor.  Error bars represent two standard deviations from the bootstrap set.\label{fig:scaling}}
\end{figure*}

\subsection{Comparison with a nearly equilibrated thermal annealer}

Hen et al.\ found strong correlation between the success probabilities of a D-Wave processor and a nearly equilibrated thermal annealer with a final inverse temperature of $\beta_f=5$.  Our results (see Fig.\ \ref{fig:scaling} and Supplemental Material \cite{supp}) show poorer correlation and an inability to fit D-Wave scaling data to SAA at a single inverse temperature.  Unlike the range-unlimited instances studied in \cite{henfl}, the hardness peak for SAA does not remain constant with varying $\beta_f$.  In the Supplemental Material we show instancewise scatter plots of D-Wave and SAA success probabilities.  The predictions of the thermal model do not correlate well with the hardware on the range 2 and range 3 instances.

\section{Discussion}

Far from being artificial, fixed coupling range in the large system limit appears, at the phase transition in the Ising formulation of Not-All-Equal 3-SAT \cite{achlioptas2001phase}\footnote{The expected number of constraints containing a given pair of variables is approximately 12.6/n at the phase transition for NAE3SAT instances on $n$ variables, whereas in a frustrated loop instance with $\alpha=0.25$ on a Chimera graph the expected number of constraints containing a given coupler is close to 1.}, which is NP-hard and can be formulated as a frustrated loop problem on the complete graph.  Furthermore, limiting coupling range in hard frustrated loop instances affects a vanishingly small portion of each instance (see Supplemental Material \cite{supp}).

The study of instances with fixed coupling range allows for the control of two factors: amplification of analog control error (for D-Wave) and effective operating temperature (for D-Wave, SAS, SAA, and any other simulated physical model with a thermal component \cite{henfl}).  Our results, taken in conjunction with those from Ref.\ \cite{henfl}, indicate a decreasing advantage for the D-Wave hardware relative to SAS with increasing range.  This observation is consistent with the hypothesis that increasing range penalizes the hardware by augmenting both the relative magnitude of control errors and the importance of thermalization.

It is straightforward to construct input classes for which analog control error will dominate the performance scaling of an analog processor.  When probing the potential for quantum speedup in a quantum annealing platform, it is important to do the opposite: construct an input class for which the impact of analog control error is minimal.  In doing so we might better observe properties of the annealer's mechanics rather than observing the effect of precision limitations, which by now are reasonably well understood \cite{Venturelli2014, King2014,henfl} and expected to improve with the maturation of the technology and the possible implementation of error correction strategies \cite{Pudenz2014,Pudenz2014a,Young2013,Vinci2015}.

\begin{acknowledgments}
The authors thank Tameem Albash, Itay Hen, Joshua Job, and Daniel Lidar for a detailed and informative exchange on this work and theirs, and for generous provision of data.  They thank Evgeny Andriyash and Jack Raymond for fruitful discussions about frustrated loops, and Emile Hoskinson for providing specifications of the D-Wave processor used.  They thank Mohammad Amin and Miles Steininger for valuable comments on the manuscript.
\end{acknowledgments}

%

\clearpage



\section*{Supplementary material}

# Supplemental material for "Performance of a quantum annealer on range-limited constraint satisfaction problems"

A. D. King,* T. Lanting, and R. Harris

*D-Wave Systems Inc., 3033 Beta Avenue, Burnaby, British Columbia, Canada V5G 4M9*
(Dated: September 3, 2015)

## I. METHODS

### A. Quantum annealing platform

In this work we used a Vesuvius quantum annealing processor manufactured by D-Wave Systems Inc. The processor is of identical design to the D-Wave Two V6 processor installed at ISI [1], and is from the same fabrication lot. We call these two processors "SR10-V6" and "ISI-V6" respectively. The problems we consider are generated on subgraphs of the processor's *Chimera* qubit connectivity graph [28], using up to 467 qubits (see Fig. 5). The data were gathered in June, 2014. SR10-V6 had an operating temperature of approximately 15 mK, and used maximum $J$ inductance (coupling strength) of 1.25 pH, compared with temperature and inductance of 17mK and 1.33pH for ISI-V6 [1]. All experiments were run using an anneal length of $t_a = 20\mu s$, equal to the runs used for most of the key analysis in the work of Hen et al. [1]. Although the processors have the same architecture and

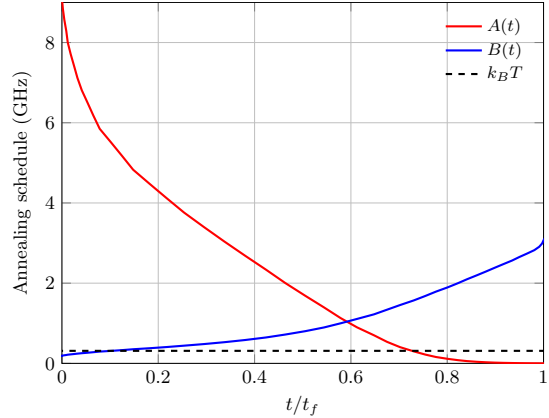

FIG. 6. **Annealing schedule of the D-Wave processor.** $A(t)$ and $B(t)$ represent the transverse field and longitudinal couplings, respectively. Operating temperature of 15mK is shown. Scale is normalized to $h = 1$.

similar ratios of temperature to inductance, they used different annealing schedules (see Fig. 6 and Ref. [1]).

### B. Experimental details

The primary testbed of problems studied consists of 200 instances of each size $L \in \{4, 5, 6, 7, 8\}$, for each value of $\alpha \in \{0.10, 0.15, \ldots, 0.50\}$, for each range limit $R \in \{2, 3, \infty\}$. Data were not collected using the hardware for $R = \infty$, as the processor was taken offline in 2014 before the need for such data was apparent. In Section II B we provide evidence that such data would likely be similar to the results in Ref. [1], where quantum annealing success probabilities are highly correlated with themselves when run on a different annealing schedule.

#### 1. D-Wave processor

Each instance was annealed 10240 times by the D-Wave processor with an anneal length of 20µs, the minimum allowed by the system. These experiments were performed in batches of 1024 anneals, each batch with a random Ising spin reversal, or *gauge transformation* applied, as in previous work [4, 32].

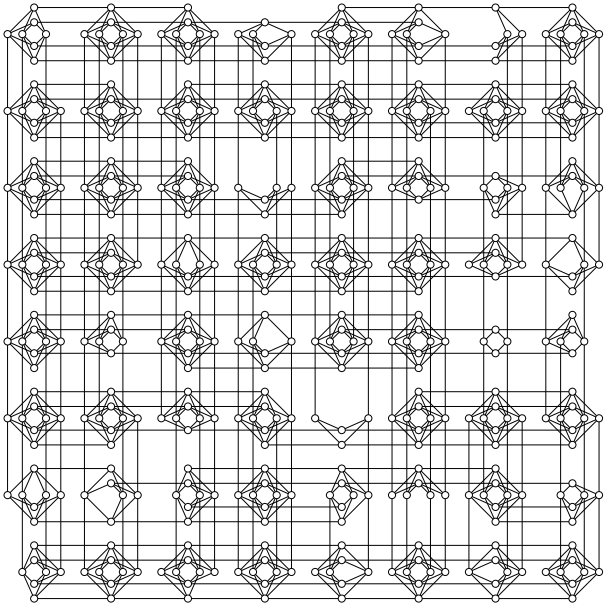

FIG. 5. The largest subprocessor used, a partial $C_8$ graph – an $8 \times 8$ grid of unit cells – with 467 qubits. Smaller instances use the square subgrid containing the top-left corner.

---

* aking@dwavesys.com

### 2. HFS

Selby's implementation of the HFS algorithm [1, 24, 25] is a heuristic approach whose main effort consists of *tree updates*, and which typically restarts to a random state when it performs two tree updates without improvement in energy. In order to treat HFS more like an annealer, we modified the code so that a solution is recorded before each reset – as in SAS, the original algorithm is modified to return possibly optimal intermediate states. It would be algorithmically equivalent to force the HFS algorithm to terminate instead of resetting.

In our experiments, each instance was solved 10240 times by the HFS algorithm. The runs were performed single-threaded in parallel using an HPC cluster of 8-core Intel Xeon E5-2670 processors.

### 3. SAS

SAS was run on each instance 1024 times at anneal length *roundoff* $(2^{a/2})$ for each integer $a \in \{2, \ldots, 22\}$. Thus the longest anneal for each instance is 2048 Monte Carlo sweeps, enough so that the optimal anneal length for median time to solution is exceeded for all problem sets studied (see Fig. 7).

### 4. SAA

SAA was run on each instance 1024 times for 20,000 Monte Carlo sweeps. For similar instances, Hen et al. support the claim that 20,000 gets us reasonably close to equilibration (of performance as a constant-time optimizer) for frustrated loop problems. We did not investigate longer SAA anneals due to limited availability of computing resources. Like SAS, SAA was run on a linear schedule in inverse temperature from $\beta_0 = 0.01$ to $\beta_f \in \{3, 4, 5\}$.

### C. Benchmarking methodology

Following previous work [1, 4, 32], we treat each solver (now including HFS) as a stochastic sampler, and measure the time to reach 99% confidence of having found the ground state energy of an instance. We call this *time to solution* (TTS). Given a solver achieving success probability $p$ over a set of trials (i.e. anneals) and taking mean time $\tau$ to complete a single trial, we compute the number of samples required as TTS for this solver and instance as

$$r(p) = \frac{\log(0.01)}{\log(1-p)}, \qquad (1)$$

and compute TTS for this solver and instances as $\tau r$. For SAS we assume $\tau$ to be 3.54μs multiplied by the number

of update sweeps in the anneal in order to remain consistent with Hen et al. [1]. We determine the optimal sweep length for each set of 200 problems as the length giving the minimum bootstrapped median time to solution. We disregard the issue of conditioning our SAS results on minimizing TTS; due to the smoothness of the curves in Fig. 7 near the optimal anneal lengths, we do not expect this to have a significant effect on the results.

For HFS our methodology differs from that of Hen et al.: We use an enumerative effort computation, as with the D-Wave processor and SAS, rather than timing the process. This allows a bare look at the dominant operations. Following Hen et al., we assume hypothetical parallelization of $L$ cores on a $C_L$ instance, and accordingly assume that a tree update on a $C_L$ instance takes $O(L)$ parallel steps (actually $L$ steps) of "leaf updates". Using these assumptions we compute the total number of leaf update steps required for a given "anneal" (i.e. sample draw) and for convenience we assume a constant of $L \cdot 1$μs per tree update, noting that this gives reasonably comparable performance at $C_4$ scale to the results of Hen et al. [1].

#### 1. Statistical methods

To generate the data points and error bars in the performance and speedup figures, we used the same Bayesian bootstrapping approach used by Hen et al. [1]. First we describe the approach for performance data, which differs slightly for HFS.

For a given solver, SR10-V6 or SAS or SAA, and each set $S$ of 200 instances at a given value of $(L, \alpha)$, we have, for each instance $s_i$, an empirical success probability $p_i$ representing $x_i$ successes out of $y$ trials. We consider the probability distribution of success probability to be $\beta_i = \beta(x_i + \frac{1}{2}, y - x_i + \frac{1}{2})$. We then construct 1000 bootstrap sets $S_j$ of size 100 by drawing 200 members from $S$ with repetition, resulting in multisets $S_j = \{s_{i,j} \mid 1 \le i \le 100\}_j$. Now for each set we sample a probability $p_{i,j}$ from distribution $\beta_{i,j}$.

At this point we can apply the desired function $f_j$ to the set of 100 probabilities, which is typically the median of $\{r(p_{i,j})\}$ for a fixed $j$. We then take the data point to be the mean of $f_j$, and we take the error in the statistic to be the standard deviation of $f_j$.

For HFS we use a similar approach suggested by Joshua Job (personal communication, January 2015). After we sample 1000 values indicating the number $r_i$ of samples needed for 99% assuredness of success, we take $\lceil r_i \rceil$ random samples (with repetition) from our set of 10240 samples, and take the sum of the numbers of tree updates in those $\lceil r_i \rceil$ samples to give a sample of time to solution.

*Speedup.* To compute data on speedup between two solvers (or equally, the same solver on two problem sets), we assume that time to solution is normally distributed with mean and standard deviation as estimated above.

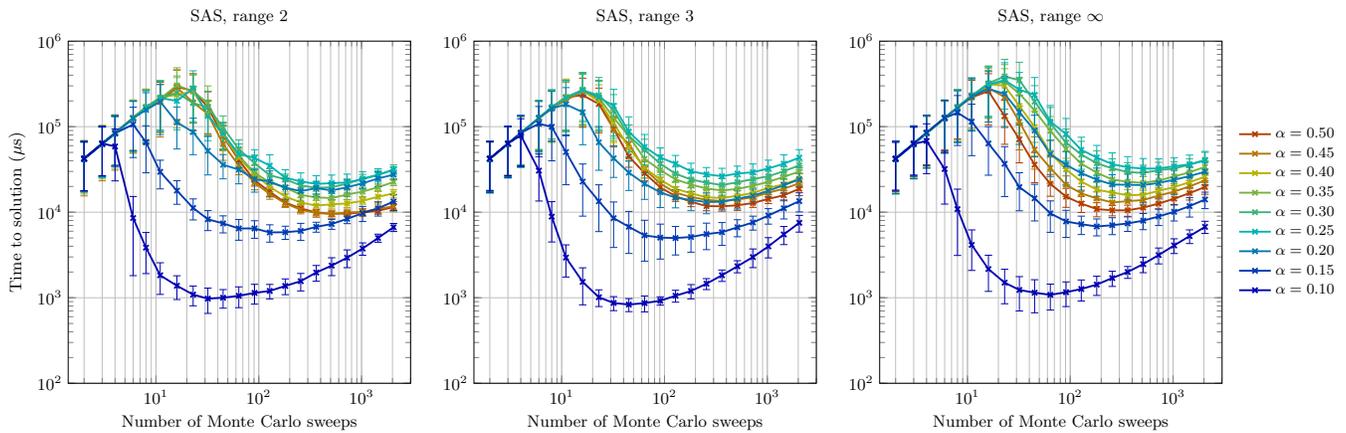

FIG. 7. **SAS time to solution vs. anneal length for $C_8$ instances.** Shown here are SAS results for $R \in \{2, 3, \infty\}$ on $C_8$ problems. Note that all panels are qualitatively similar. Behavior at the left of each panel is a statistical artifact arising from our Bayesian model and limited number of experiments, which effectively places a lower bound on the success probability (cf. the data of Hen et al. on suboptimal annealing time [1]). The slight increase in difficulty for the hardest problems as $R$ increases can be explained by higher temperature relative to the gap of the final Hamiltonian. The increase in difficulty for long anneals in high-$\alpha$ problems as $R$ increases can be explained by the suppression of randomness (and accelerated evolution of ferromagnetic structure) for low-$R$, high-$\alpha$ combinations (see Appendix II B).

We then sample 1000 points from each normal distribution and compute 1000 speedup ratios. We use the mean and standard deviation of this set of ratios for our data point and error bar.

*Scaling coefficients.* When computing scaling coefficients (see Appendix II C), we assume normal distributions on TTS for a given solver and set of 1000 instances (200 of each size). We then draw 1000 samples from each of the five distributions, and for each set of five samples we compute the slope of the best fit line $\ln(\text{TTS}) \approx a(\alpha) + b(\alpha)L$. From these 1000 slopes we take the mean and standard deviation. Error bars in Fig. 3 represent two standard deviations, in keeping close to the methods of Hen et al., who use 95% confidence intervals [1].

## II. FURTHER DATA

### A. Problem hardness as a function of constraint-to-variable ratio

In Section III we described Itay Hen's original construction of frustrated loop instances, and offered a modification. For convenience, we call the former *HenFL* instances, and the latter *KingFL* instances. As explored by Hen et al. [1], there is a clear easy-hard-easy pattern of hardness for all solvers as $\alpha$ increases from 0 to 1 and beyond, with the hardest regime generally falling near the point $\alpha = 0.25$. This is shown to correlate with frustration in the problem, measured by Hen et al. as the proportion of extant couplers that are frustrated in the planted ground state. For large $\alpha$, systems tend towards ferromagnetism. For sufficiently small $\alpha$, a typical system is simply a collection of disjoint frustrated loops, and therefore both highly degenerate and combinatorially trivial. In Fig. 8 we see the same qualitative dependence on $\alpha$.

In HenFL problems, the mean loop length is approximately 11. In KingFL problems it is approximately 9 (see Fig. 9 for the instances studied). It is therefore not surprising that the hardest problems appear for slightly larger $\alpha$ in our results compared with the results of Hen et al. We remark that a well-yielded Chimera graph will have between $2n/5$ and $3n$ edges, so in our data the hardest problems arise when the expected number of loops containing a given coupler is near 1.

### B. Effect of range limitation and loop distribution

Of fundamental importance to this work is the question of whether or not limiting the coupling range of frustrated loop instances makes them intrinsically easier. Here we give evidence that the difference in performance of the D-Wave processors here and in the paper of Hen et al. [1] cannot be explained by our testbed being easier.

Clearly the range-limited instances studied here are easier for the SR10-V6 D-Wave processor used in this work than the range-unlimited HenFL instances are for the ISI-V6 processor, and clearly the range-2 instances are easier than the range-3 instances for SR10-V6. For an idea of whether or not range-limited instances are *combinatorially* easier, we appeal to HFS performance, since HFS is unaffected by coupling range and has no thermal component. Perhaps the most pertinent answer to this question is in the bottom-middle panel of Fig. 11. There we see that for the highest $\alpha$, HFS finds range-

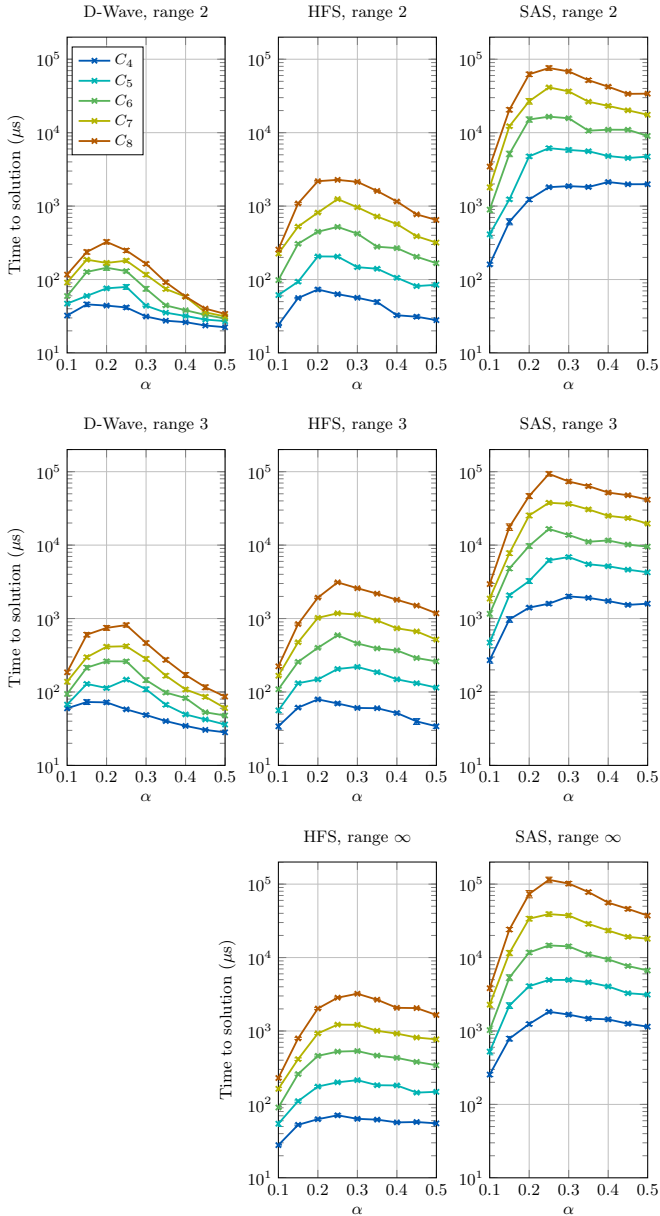

FIG. 8. **Time to solution vs. qubit-to-constraint ratio** $\alpha$. Shown is the median time to solution for the D-Wave processor (left), HFS (middle), and SAS (right) plotted against $\alpha$ for each problem size $C_L$ for $L \in \{4, 5, 6, 7, 8\}$.

2 instances consistently easier than range-$\infty$ instances, but this phenomenon weakens as $\alpha$ decreases. There is a simple intuitive reason for this: for any given coupling range limit $R$, if $\alpha$ is sufficiently large (but not so large that a random KingFL instance cannot be consistently constructed), the number of loops containing an edge is forced to be relatively consistent across the edges of the graph. Consequently, the system is more orderly and therefore more ferromagnetic – recall that at least 5/6 of nonzero couplers in each $J_i$ are ferromagnetic. This intuition is corroborated by our HFS results.

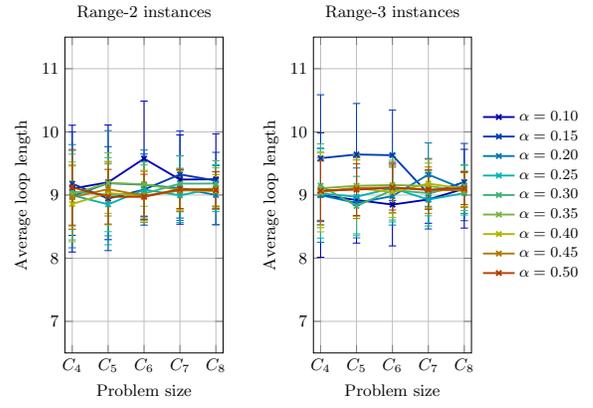

FIG. 9. **Loop length in experimental testbed.** Shown here are the mean and standard deviation of loop length for range-2 and range-3 instances. The distribution appears to converge to approximately length 9. This may be slightly different for more fully-yielded Chimera graphs.

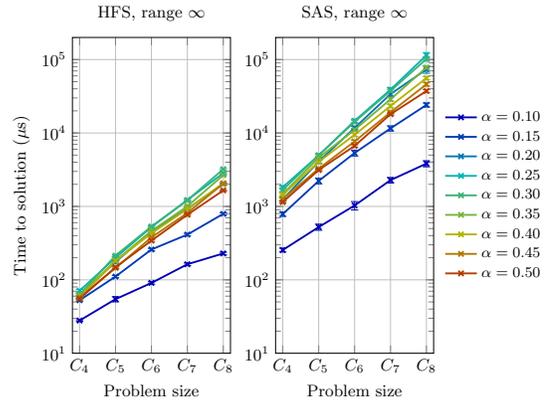

FIG. 10. **Median time to solution per size for range-unlimited instances.** Shown here are data for HFS and SAS on range-unlimited instances; these plots are analogous to Fig. 1.

In Figures 12 and 13 we give HFS data on a more extensive set of inputs, arranged according to $\alpha$. For each choice of $L$ and $\alpha$, the set contains 200 instances. Fig. 12 suggests that we should not expect the range-unlimited KingFL instances on the SR10-V6 hardware graph to be any easier than the range-unlimited HenFL instances on the ISI-V6 hardware graph, if for each set we choose $\alpha$ to maximize median HFS time to solution. This takes into consideration the fact that ISI-V6 uses 503 qubits on $C_8$ problems, while SR10-V6 uses only 467. Fig. 13 suggests that for $L \leq 8$ and $\alpha \leq 0.3$, we should not expect range-3 KingFL instances to be consistently easier than range-unlimited KingFL instances. For $L \leq 8$ and $\alpha \leq 0.2$, we should not expect range-2 KingFL instances to be consistently easier than range-unlimited KingFL instances. In short, it appears that the difference between our results and the results of Hen et al. [1], especially the scaling advantage we see for the D-Wave processor on

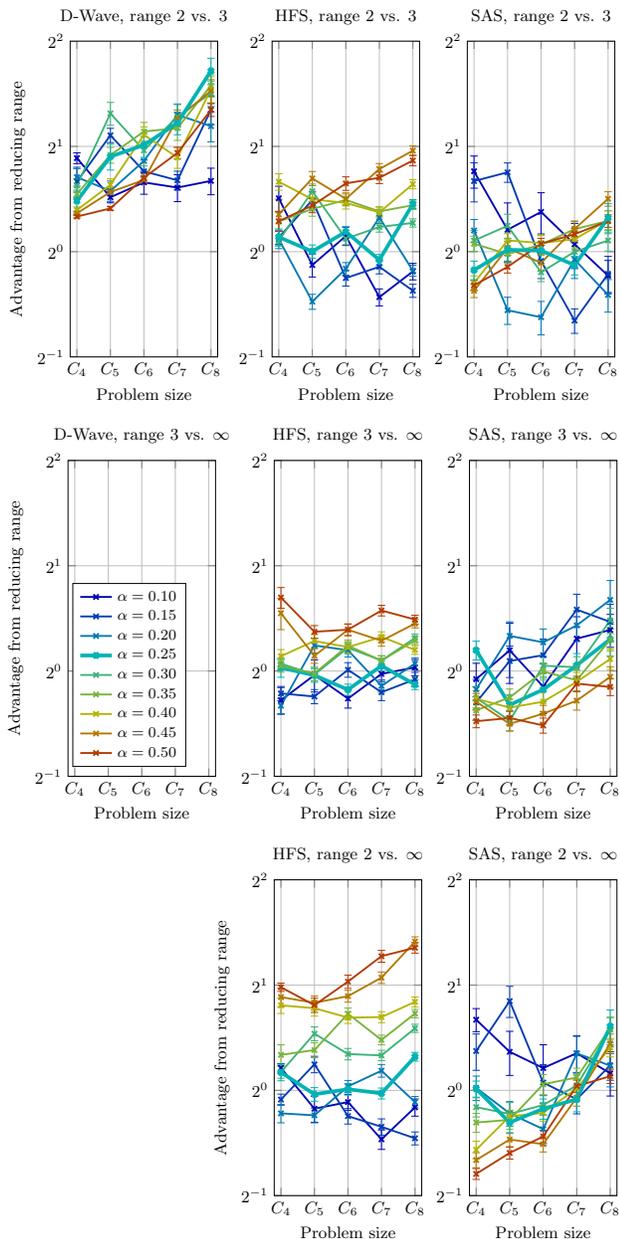

FIG. 11. **Benefit of reduced coupling range.** Here we show the median speedup arising from changing the coupling range for each solver. No range-$\infty$ data exists for the D-Wave processor. Positive slopes indicate increasing benefit from reduction of precision range as problems get larger. Error bars represent one standard deviation from bootstrap samples.

range-3 instances, cannot be attributed to the problems being fundamentally easier.

For SAS, the issue of computational difficulty is convolved with that of effective temperature. This is most troublesome for range-$\infty$ instances, where the energy scale (and relative temperature) varies significantly between instances (see Fig. 14). Speedup for range-3 over range-2 problems for low $\alpha$ may indicate that the final

inverse temperature $\beta_f$ is too low relative to the energy scale of the normalized Hamiltonian, whereas speedup for range-3 over range-2 problems for high $\alpha$ may indicate, in agreement with HFS data, that these problems are easier in range 2. Fig. 7 provides further insight into the question of easiness and temperature.

Going back to Fig. 11, we emphasize that particularly for $\alpha \leq 0.3$, which includes the hardest instances, the HFS speedup as coupling range decreases is small compared with both the speedup of the D-Wave processor over HFS and the speedup of the D-Wave processor between range-2 and range-3 instances. For $\alpha \leq 0.25$, range-3 instances appear to have very little structural "easiness" compared with range-$\infty$ instances.

As further illustration of why this should be the case, Fig. 15 takes the range-$\infty$ KingFL testbed and plots the proportion of nonzero couplers exceeding a given range limit for each value of $\alpha$. For a fixed $\alpha$, the structural impact of imposing a coupler range limit of $R$ diminishes quickly as $R$ increases. In other words, in range-$\infty$ instances scaled to the interval $[-1, 1]$, the scaling factor is determined by the tail of the coupler value distribution, which we expect to be insignificant to the overall combinatorial structure of the problem.

### C. Ruling out a thermal model for D-Wave performance

In order to support a thermal model for performance scaling of the D-Wave processor, SAA data should provide a good match with D-Wave processor data on both range-2 and range-3 instances at the same final inverse temperature $\beta_f$, for various choices of $\alpha$. In other words, it should have predictive power [11]. Hen et al. found a good match between SAA and their D-Wave processor data at $\beta_f = 5$ for certain values of $\alpha$ [1]. In Fig. 3 we can see the scaling coefficients for the D-Wave processor and for SAA with $\beta_f \in \{3, 4, 5\}$. The ratio of temperature to energy scale is similar for the D-Wave processor studied here and that studied by Hen et al. ($\sim 7\%$ difference). For range-3 instances, $\beta_f = 4$ appears to be too high and $\beta_f = 3$ is clearly too low. Although $\beta_f = 4$ gives a better fit, its scaling does not match that of the D-Wave processor on low-$\alpha$ range-2 instances.

Fig. 16 shows instance-wise scatter plots of success probabilities for the D-Wave processor and SAA at $\beta_f = 4$. It is clear from this data that the best correlation is poor when compared with the data of Hen et al., particularly looking at each problem size individually, and that there is not a good fit between success probabilities on range-2 instances. These data fail to support the hypothesis that SAA provides a thermal model for performance scaling on frustrated loop instances. On the contrary, there are several pieces of evidence that in this context, analog control error causes the appearance of thermal behavior. First is the poor correlation compared to that found by Hen et al.; the data shown in Fig. 16 for

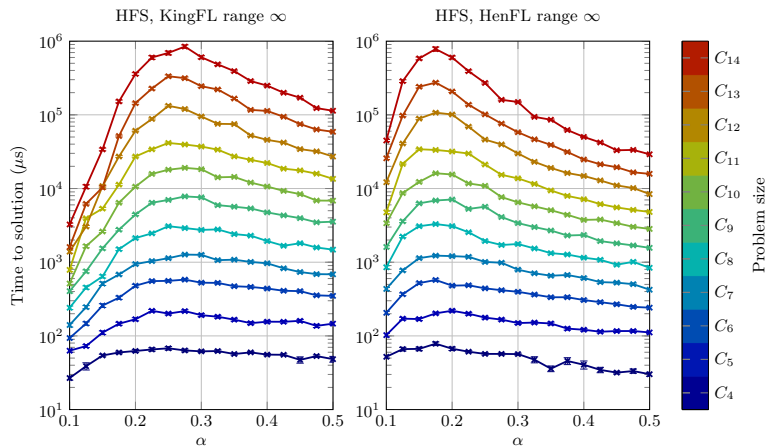

FIG. 12. **Comparison of KingFL and HenFL instances on different hardware graphs.** Shown here are median TTS for HFS on range-unlimited KingFL and HenFL instances. The KingFL instances are constructed on subgrids of a random $C_{16}$ graph with similar yield to the SR10-V6 processor (91%), while HenFL instances are constructed on subgrids of a random $C_{16}$ graph with similar yield to the ISI-V6 processor (98%).

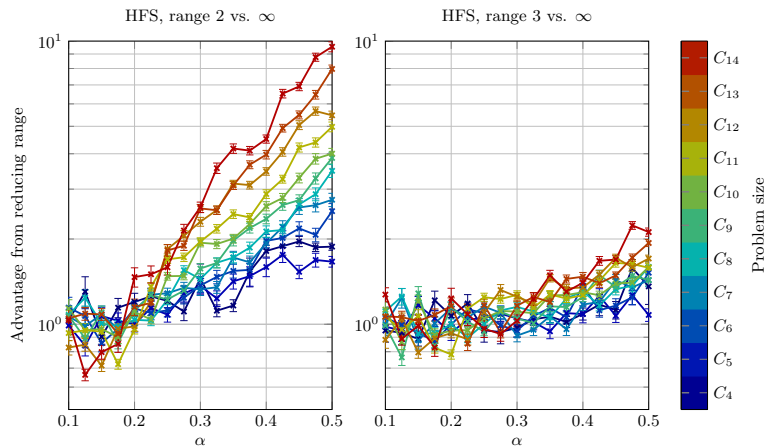

FIG. 13. **Benefit of reduced coupling range for HFS.** Here we show median speedup arising from changing the coupling range for HFS on a broad set of inputs: $\alpha \in \{0.100, 0.125, \ldots, 0.500\}$ and $L \in \{4, \ldots, 14\}$. Instances are constructed on full $C_L$ graphs. Error bars represent one standard deviation from bootstrap samples.

range-3 instances represents the best visual fit between the D-Wave processor and SAA data out of all our experiments. Second is the fact that we do not find agreement at the same inverse temperature $\beta_f = 5$. Third is given by Hen et al. ([1] Fig. 12), who show scaling coefficients for various choices of $\beta_f$, along with results on perturbed instances at $\beta_f = 5$. Their data suggests that perturbing the Hamiltonian has a similar effect to increasing the apparent temperature of SAA.

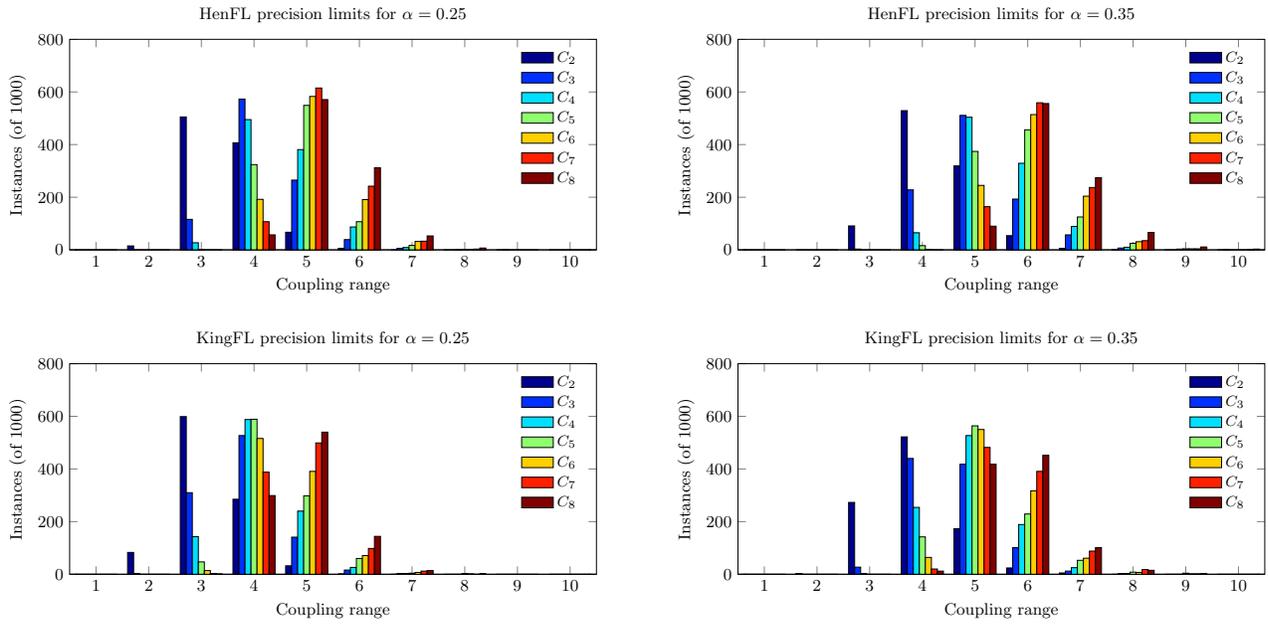

FIG. 14. **Distribution of coupling range for range-unlimited instances.** Here we see the change in the distribution of coupling range for random HenFL and KingFL instances at $\alpha = 0.25$ and $\alpha = 0.35$. Each data series consists of 1000 randomly generated instances on a full Chimera graph $C_L$. Naturally, coupling range increases as $\alpha$ increases. KingFL instances have lower coupling range than corresponding HenFL instances, owing to their lower average loop length.

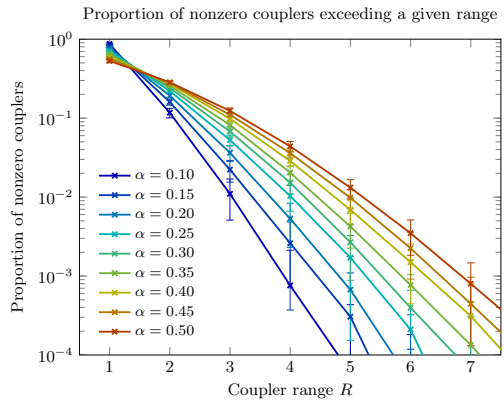

FIG. 15. **Couplers exceeding a given range.** Shown is the mean proportion of nonzero couplers in a $C_8$ range-$\infty$ exceeding each range limit. For fixed $\alpha$, the proportion of couplers exceeding the range limit $R$ decreases superexponentially with $R$. Error bars represent one standard deviation from each data set of 200 instances.

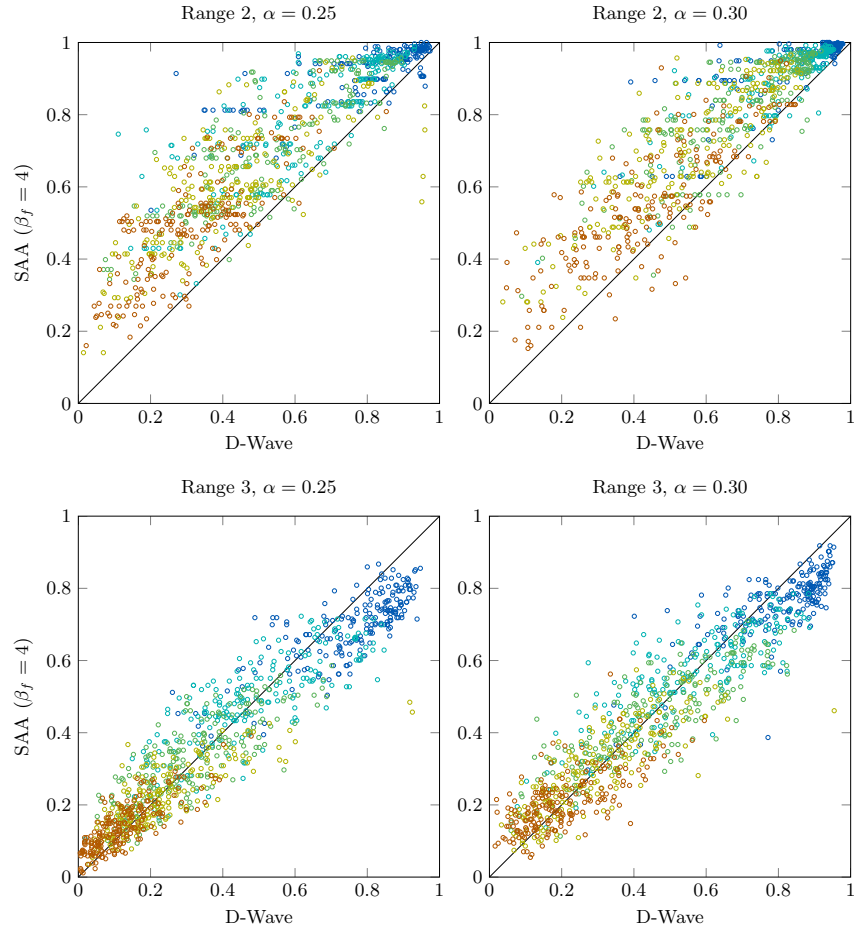

FIG. 16. **Success probability correlations between the D-Wave processor and SAA.** Shown are instance-wise scatters for the most difficult problems ($\alpha \in \{0.25, 0.30\}$) for range-2 and range-3 instances. Instances receive colors according to their size. SAA has $\beta_f = 4$. Correlation on the range-3 instances is weak compared to correlation found by Hen et al. [1], whereas correlation on range-2 instances, particularly for each problem size as a separate data set, is poor.


\end{document}